\documentclass[aps,pre,floatfix,superscriptaddress,notitlepage,longbibliography]{revtex4-1}
\usepackage{graphicx}
\usepackage{epsfig}
\usepackage[normalem]{ulem}
\usepackage{subfigure}

\newcommand{\be}{\begin{equation}}
\newcommand{\ee}{\end{equation}}

\usepackage{amsmath}

\usepackage{bm} 
\usepackage{amssymb}
\usepackage{natbib}
\usepackage{graphicx}
\usepackage{color}

\begin{document}

\title{Three-dimensional active turbulence in microswimmer suspensions: simulations and modelling.}

\author{A. Gasc\'o}
\affiliation{Departament de F\'{\i}sica de la Mat\`eria Condensada, Universitat de Barcelona, Carrer de Mart\'{\i} i Franqu\`es, 08028 Barcelona, Spain}
\author{I. Pagonabarraga}
\affiliation{Departament de F\'{\i}sica de la Mat\`eria Condensada, Universitat de Barcelona, Carrer de Mart\'{\i} i Franqu\`es, 08028 Barcelona, Spain}%
\affiliation{  
Universitat de Barcelona Institute of Complex Systems (UBICS), Universitat de Barcelona, 08028 Barcelona, Spain}
\author{A. Scagliarini}
\email[]{andrea.scagliarini@cnr.it}
\affiliation{Istituto per le Applicazioni del Calcolo (IAC), Consiglio Nazionale delle Ricerche (CNR), Via dei Taurini 19, 00185 Rome, Italy}
\affiliation{INFN, sezione Roma ``Tor Vergata'', via della Ricerca Scientifica 1, 00133 Rome, Italy}

\begin{abstract}
  \noindent Active turbulence is a paradigmatic and fascinating example of self-organized motion at
  large scales occurring in active matter.
  We employ massive hydrodynamic simulations of suspensions of resolved model 
  microswimmers to tackle the phenomenon in semi-dilute conditions at a 
  mesoscopic level. We measure the kinetic energy spectrum and 
  we detect a $k^{-3}$ power law regime. The velocity distributions are of L\'evy type, 
  a distinct difference with inertial turbulence.
  Furthermore, we propose a reduced order dynamical deterministic model for active
  turbulence, inspired to {\it shell models} for classical turbulence, whose 
  numerical and analytical study confirms the spectrum power-law observed in the simulations
  and reveals hints of a non-Gaussian, intermittent, physics of active turbulence.
  Direct numerical simulations and modelling also agree in pointing to a 
  phenomenological picture whereby, in the absence of an energy cascade 
  {\it \`a la Richardson} forbidden by the low Reynolds number regime, it is the 
  coupling between fluid velocity gradients and bacterial orientation 
  that gives rise to a multiscale dynamics.

\end{abstract}  

\maketitle

\section{Introduction}
\noindent One of the most striking features of active systems is the emergence of correlated motion and
structures on scales much larger than that of the single agents. Forms of self-organized motion
are ubiquitous in Nature~\cite{VicsekPR2012},
from colonies of microorganisms~\cite{KearnsNature2010,DarntonBJ2010,KochARFM2011}
to flocks of birds and fish schools~\cite{BalleriniPNAS2008,HerbertPNAS2011}.
Collective motion in microbial suspensions, in particular, appears as a tangle of coherent structures,
like vortices, jets, that recall those typically encountered in turbulent flows.
This morphological analogy suggested to introduce the term {\it bacterial} or
{\it active turbulence}~\cite{DombrowskiPRL2004,WolgemuthBJ2008,PedleyARFM1992,AlertARCMP2022}. 
But {\it what is turbulence}? Borrowing a celebrated quote, it is 
"hard to define, but easy to recognize when we see it!"~\cite{Vallis}. 
Is, then, visual inspection enough? Certainly not, as witnessed by the consistent 
body of works devoted to a quantitative description of 
complex motion in active systems and of its relation with turbulence. 
Power-law decays of kinetic energy spectra over unexpectedly wide ranges of wavenumbers have been identified as a signature of 
turbulent behaviour in diverse systems including dense bacterial suspensions~\cite{IshikawaPRL2011,WensinkPNAS2012}, collections of 
ferromagnetic spinners~\cite{KokotPNAS2017}, 
active nematics~\cite{DoostmohammadiNATCOMM2017,sagues2021}, 
synthetic swimmers at interfaces~\cite{BourgoinPRX2020}, among others.
Nevertheless, a {\it universality} of some kind seems to be 
lacking~\cite{WensinkPNAS2012,BratanovPNAS2015}, unlike in classical turbulence, where Kolmogorov's theory of the 
inertial range constitutes a key, unifying ingredient~\cite{Frisch}. With the aim of analyzing 
multiscale interactions and energy transfer, recent studies have focused on the 
spectral properties of continuum
models~\cite{SlomkaPNAS2017,LinkmannPRL2019,CarenzaPRF2020,CarenzaEPL2020,AlertNP2020,RoraiPRL2022},
where the active fluid is described as an effective medium, whose equations
are inspired by 
nematodynamics~\cite{KrusePRL2004,GiomiPRL2008,EdwardsEPL2009,TjhungSM2011} and pattern formation~\cite{WensinkPNAS2012,SlomkaEPJST2015}.
In all such approaches, it is surmised that the modelled system, be it 
a bacterial suspension or an active gel, is highly concentrated, such that 
the interactions among active agents are essentially excluded volume 
and alignment.
On the other hand, in the limit of extreme dilution (i.e. for volume fractions 
below the onset of collective motion), it has been shown theoretically that, due to long-range hydrodynamic interactions, suspensions of 
extensile microswimmers ("pushers") develop a solvent velocity field with variance 
anomalously growing with the volume
fraction and kinetic energy spectra whose functional form can be derived analytically with a kinetic theory approach~\cite{StenhammarPRL2017,BardfalvySM2019,SkultetyPRX2020}.
In semi-dilute conditions, hydrodynamically induced correlations engender
collective motion and large-scale flows~\cite{WuPRL2000,HernandezOrtizPRL2005,LeptosPRL2009,SaintillianJRSI2012}.
In this context, power-law spectra have been measured
in numerical simulations of suspensions of point-like and slender rod-like 
microswimmers~\cite{SaintillianJRSI2012,BardfalvySM2019}.
However, despite these few relevant exceptions, the phenomenon of active turbulence in 
semi-dilute situations, namely far from close packing but above the onset of 
collective motion,
has been much less investigated and is still poorly understood from a theoretical point of view.
So, the goal of the present paper is twofold. On one hand, we will describe active turbulence signatures in semi-dilute suspensions of model spherical swimmers (squirmers); on the other 
we will propose a novel theoretical framework helping to understand the mechanisms underlying active 
turbulence phenomenology.
\\
To this aim, we perform large scale direct numerical simulations (DNS) of suspensions of pushers, 
where the solvent hydrodynamics is fully resolved, both
from the near to the far field. 
We characterize the solvent velocity field in terms of its energy spectra
and probability density functions (PDFs). The spectrum displays a {\it plateau} at small wavenumbers, consistently with theoretical predictions~\cite{BardfalvySM2019,SkultetyPRX2020} for very dilute systems, followed by an algebraic decay, $k^{-3}$, signalling the presence of fluid motion on scales 
up to roughly half a decade larger than the microswimmer's size.\\
We also introduce a reduced order dynamical deterministic model of 
active turbulence, 
pertaining to the class of the so called {\it shell models}, motivated by 
their successful story as turbulence models~\cite{BiferaleARFM2003}.  
Having at disposal a shell model of active turbulence 
allows to study the statistical properties and chaotic behaviour of a 
computationally and theoretically challenging phenomenon 
within a low number of degrees-of-freedom description, which is, to some extent, 
even amenable of analytical treatment, as we also prove.
The analysis of the shell model reproduces the kinetic energy spectrum power-law observed in the DNS.
The shape of the velocity variables PDFs break scale-invariance, a hallmark of intermittency.
The shell model leverages a phenomenological picture, confirmed by the DNS, 
whereby the coupling between the fluid velocity gradients and the bacterial orientation 
dynamics is the mechanism responsible for the generation of flow at large scales.

\section{Methods}
\noindent The solvent hydrodynamics is simulated using a standard $D3Q19$
lattice Boltzmann method~\cite{Succi,WolfGladrow,DesplatCPC2001}. 
The microswimmers are modelled as solid spheres of radius $R$, 
mass $M=\frac{4}{3}\pi R^3\rho_p$ (where $\rho_p$ is the microswimmer density) and moment of inertia 
$I=\frac{2}{5}MR^2$.
The momentum/torque exchange in fluid-solid coupling is ensured by the so called bounce-back-on-links
algorithm~\cite{LaddJFM1994,NguyenPRE2002,AidunARFM2010}. 
The swimming mechanism is introduced via a minimal implementation of the ``squirmer'' model~\cite{BlakeJFM1971,IshikawaJFM2006}, whereby
a non-zero tangential polar component of the axisymmetric slip velocity profile, 
depending on the two parameters $B_1$ and $B_2$, is imposed
at the particle surface 
\begin{equation}\label{eq:uslip}
\mathbf{u}_s = (B_1 \sin\theta + B_2 \sin \theta \cos \theta)\hat{\mathbf{\theta}},
\end{equation}
where $\theta = \arccos(\hat{\mathbf{e}}\cdot \hat{\mathbf{r}}_s)$ is the angle between the 
squirmer orientation unit vector, $\hat{\mathbf{e}}$, and the position on the 
particle surface, $\hat{\mathbf{r}}_s= \frac{\mathbf{x}_s - \mathbf{X}}{R}$, relative 
to the position of its centre of mass, $\mathbf{X}$ (i.e. it is the polar angle for a spherical coordinate
system in a frame comoving with the particle and $\hat{\theta}$ is its associated unit vector).
With the prescription (\ref{eq:uslip}) for the surface slip, the squirmer generates a velocity field at the position $\mathbf{x}$ that,
to leading orders, takes the form

\begin{equation}\label{eq:Ufield}
\mathbf{U}(\mathbf{x},t) = B_1\left(\left(\frac{R}{r}\right)^3(\hat{\mathbf{e}}\cdot\hat{\mathbf{r}})\hat{\mathbf{r}}-\frac{1}{3}\left(\frac{R}{r}\right)^3 \hat{\mathbf{e}} -  \frac{1}{2}\beta\left(\frac{R}{r}\right)^2\left(3(\hat{\mathbf{e}}\cdot\hat{\mathbf{r}})^2-1\right)\hat{\mathbf{r}}\right) 
+ O\left(\left(\frac{R}{r}\right)^4\right)
\end{equation}
where $\mathbf{r}=\mathbf{x}-\mathbf{X}(t)$, $r=|\mathbf{r}|$, $\hat{\mathbf{r}} = \frac{\mathbf{r}}{r}$
and $\beta \equiv B_2/B_1$ ($\beta<0$ since we are considering here pushers) determines the amplitude 
of the stress exerted by the microswimmer on the surrounding fluid.
The equations of motion for the centre of mass position
  $\mathbf{X}^{(i)}(t)$, barycentric velocity $\mathbf{V}^{(i)}(t)$, intrinsic orientation
  $\hat{\mathbf{e}}^{(i)}(t)$ and angular velocity $\mathbf{\Omega}^{(i)}(t)$ of the $i$-th
  squirmer ($i=1,2,\dots,N$) read:
\begin{eqnarray} \label{eq:eom2}
\dot{\mathbf{X}}^{(i)} &=& \mathbf{V}^{(i)} \\ \nonumber
\dot{\mathbf{V}}^{(i)}  &=& \frac{1}{M}\left(\mathbf{F}_h + \mathbf{F}_a
                            -\chi \mathbf{V}^{(i)} \right) \\ \nonumber
\dot{\hat{\mathbf{e}}}^{(i)}  &=& \mathbf{\Omega}^{(i)} \wedge \hat{\mathbf{e}}^{(i)}   \\ \nonumber
\dot{\mathbf{\Omega}}^{(i)} &=& \frac{1}{I}\left(\mathbf{T}_h -\zeta \mathbf{\Omega}^{(i)}\right),
\end{eqnarray}  
where $\mathbf{F}_a \propto \int\int_{\mathcal{S}} \mathbf{u}_s d\mathbf{x}_s$ is the force determining 
the self-propulsion (an equal in magnitude and opposite sign is imparted to the fluid such that, globally, 
the system is force-free), $\mathbf{F}_h$ and $\mathbf{T}_h$ are force and torque arising from the flows
generated in the solvent (and responsible, therefore, for hydrodynamic interactions among particles) and 
$\chi$ and $\zeta$ are friction coefficients. Equations (\ref{eq:eom2}) are solved numerically, 
time-marching first the positions and orientations vector by means of a forward-Euler
scheme, and then the velocities and angular velocities by
means of an implicit (backward-Euler) update.
The over-damped limit of (\ref{eq:eom2}) yields 
$\mathbf{V}^{(i)} \propto \Gamma^{-1} (\mathbf{F}_h + \mathbf{F}_a)$; in particular, for an isolated 
squirmer, this endows the particle with a self-propulsion velocity $V_p = \frac{2}{3}B_1 \hat{\mathbf{e}}$.
The method has been extensively tested and applied to various physical problems, including, among others, pairwise hydrodynamic interactions, the formation of 
polar order, clustering and sedimentation 
in suspensions of microswimmers~\cite{MatasNavarroEPJE2010,AlarconJML2013,AlarconSM2017,ScagliariniSM2022}.\\
We perform numerical simulations in a triperiodic cubic box of side $L=512$ lattice points, with 
$N \approx 2.6 \times 10^5$ pushers of radius
$R=2.3$ lattice units (corresponding to a volume fraction of $\phi \approx 0.1$) and $\beta=-5$ 
(for this value of $\beta$ the system does not develop a global polar order~\cite{EvansPOF2011,AlarconJML2013}). 
The first squirming parameter is set to $B_1 = 1.5 \times 10^{-3}$, so the intrinsic self-propulsion speed is $V_p = 10^{-3}$,
in lattice Boltzmann units (lbu).
The kinematic viscosity is $\nu=1/6$ lbu such that the Reynolds number at the particle scale, for an isolated microswimmer, is $\text{Re}_p \approx 10^{-2}$.

\section{Results}
\subsection{Numerical simulations}
\begin{figure}
\begin{center}
  \advance\leftskip-0.55cm
  \includegraphics[scale=0.8]{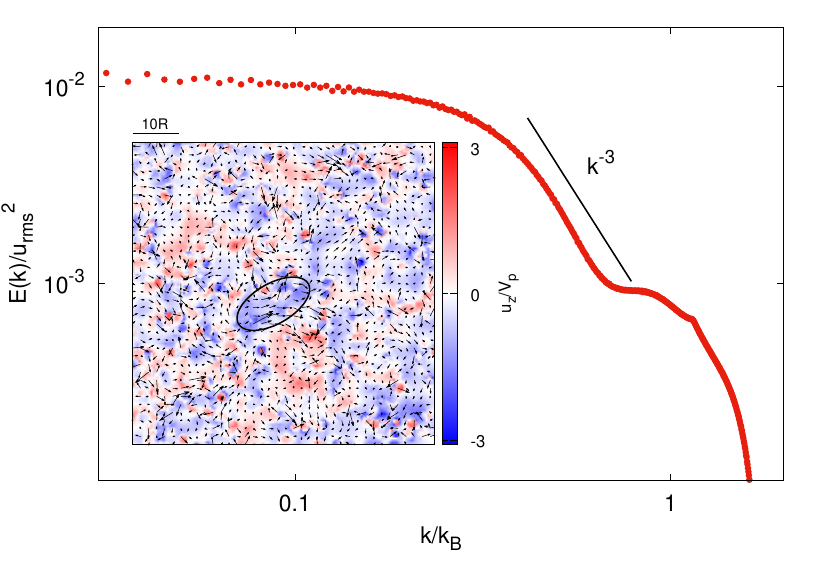}
  \caption{MAIN PANEL: Energy spectrum (time-averaged over the statistically stationary state) of the fluid velocity field, normalized by the mean square velocity $u_{\text{rms}}^2$, in a 
  suspension of pushers ($\beta=-5$) at a volume fraction $\phi \approx 0.1$ (wavenumbers are normalized by $k_B = 2\pi/R$); the solid line indicates
    the scaling $E(k) \sim k^{-3}$. INSET: Snapshot of the velocity field in the plane $z=L/2$ from a simulation: the in-plane vectors $(u_x,u_y)$ are depicted as arrows and the out-of-plane component $u_z$ (rescaled by the characteristic swimming speed $V_p$) as a color map. Notice the size
    of correlated regions as compared to the microswimmer size $\sim R$; in particular, the occurrence of a {\it jet} extending on a scale of several particle radii is highlighted by the \textcolor{red}{black} ellipse.}
\label{fig:spectrum-dns}
\end{center}
\end{figure}

\noindent In Fig.~\ref{fig:spectrum-dns} we report the energy spectrum
$E(k)=\frac{1}{2} \overline{\langle \tilde{\mathbf{u}}^{\ast}\cdot\tilde{\mathbf{u}} \rangle}$,
where $\tilde{\mathbf{u}}(\mathbf{k})$ is the Fourier transform of the fluid velocity field and the overlined brackets, 
$\overline{\langle (\cdot)\rangle}$, indicate a
surface integral on spheres of radius $k$, in spectral space, and time averaging over the statistically stationary state
(wavenumbers are normalized by $k_B = 2\pi/R$).
The spectrum shows a {\it plateau}, $E(k) \sim \text{const}$, at small wavenumbers, followed by an algebraic decay, $E(k) \sim k^{-3}$. 
This decay covers a relatively short range of wavenumbers. We must stress, though, that scaling ranges of limited extension (typically less than a decade) are a distinctive feature of a large variety of active systems \cite{WensinkPNAS2012,CarenzaPRF2020,LiuSM2021,sagues2021,lin2021}.
A constant spectrum is the theoretical expectation for an ideal system of weakly interacting point-like stresslets~\cite{StenhammarPRL2017,BardfalvySM2019,SkultetyPRX2020}, i.e. in the limit of extreme dilution.
As the concentration increases, non-linearities stemming from hydrodynamic interactions
  among particles are expected to determine the development of a power-law profile of the spectrum, with a slope that depends
  on the volume fraction~\cite{SaintillianJRSI2012,BardfalvySM2019,LiuSM2021}.
  Remarkably, though, as shown in recent experiments with {\it E. Coli},
  the exponent tends to approach the limiting value of $\sim -3$ as the systems enters
  the active turbulence regime~\cite{LiuSM2021}. It might be expected, despite the
  lack of a formal proof, that such an asymptotic behaviour holds only under semi-dilute
  conditions, i.e. below the close packing volume fraction, above which the physics of collective
  behaviour hinges on different mechanisms.
\begin{figure}
\begin{center}
  \advance\leftskip-0.55cm
  \includegraphics[scale=0.8]{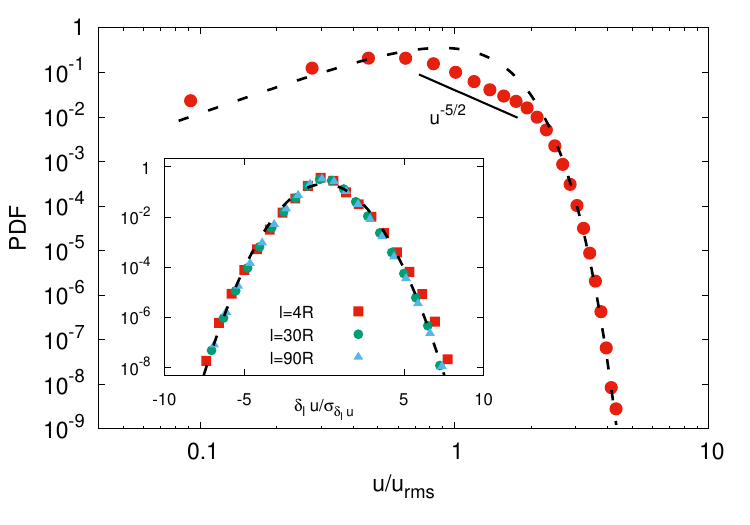}
  \caption{MAIN PANEL. Probability density function of the fluid velocity 
  magnitude (normalized by its root mean square value);
  the dashed line represents the $\chi$-like distribution $\propto u^2 e^{-\frac{u^2}{2\sigma^2}}$ (with 
  $\sigma \approx 0.63$), whereas the solid line highlights the algebraic decay $\propto u^{-5/2}$.
  INSET. Probability density functions of the longitudinal velocity increments, 
    $\delta_l u = \left[\mathbf{u}(\mathbf{x}+\mathbf{l},t) - \mathbf{u}(\mathbf{x},t)\right]\cdot\frac{\mathbf{l}}{l}$ 
  (divided by their standard deviation, $\sigma_{\delta_l u}$),
    for three separations $l/R=4, 30, 90$. 
    The dashed line depicts the Gaussian probability density
    function with zero mean and standard deviation $\sigma \approx 1.27$.}
\label{fig:pdfs-dns}
\end{center}
\end{figure}
Another peculiar aspect of the complex non-linear phenomenology of inertial turbulence is
intermittency, which is intimately related to the breakup of global scale invariance~\cite{Frisch}. 
Intermittency can be detected in 3D turbulent velocity fields looking at the probability density functions (PDFs) of longitudinal velocity increments 
$\delta_l u = \left[\mathbf{u}(\mathbf{x}+\mathbf{l},t) - 
\mathbf{u}(\mathbf{x},t)\right]\cdot\frac{\mathbf{l}}{l}$.
Upon proper rescaling, such to have a fixed variance, the PDFs do not attain a scale-independent functional form:
they are Gaussian when the separation is of the order of the system size, $l \sim L$ (the so called ``integral scale''), but become more and more {\it fat-tailed} at
decreasing $l$ (or, equivalently, the {\it flatness} grows as $l \rightarrow 0$).
In contrast, in our simulations, as shown in Fig.~\ref{fig:pdfs-dns} (inset), the PDFs seem to collapse onto a Gaussian curve for all $l$.
As for the statistics of fluid velocity, we observe from the main panel of Fig.~\ref{fig:pdfs-dns} that the PDF is well fitted by a $\chi$-like distribution, i.e. it follows a Gaussian statistics, for
small ($u \ll u_{\text{rms}}$) and large ($u \gg u_{\text{rms}}$) velocities, consistently with 
theoretical results obtained for suspensions of swimmers generating algebraically decaying flow fields, with a short range regularization~\cite{ZaidJRSI2011}. At intermediate values the PDF is expected to deviate 
from Gaussianity; here, in particular, we find a good fit by a power law with the exponent $-5/2$, which
is what would stem from a superposition of swimmers uniformly distributed in space, with associated 
velocity fields decaying as $U(r) \sim u^{-2}$~\cite{RushkinPRL2010}.
 
\subsection{A reduced order deterministic model of active turbulence} 
\begin{figure}
\begin{center}
  \advance\leftskip-0.55cm
  \includegraphics[scale=0.8]{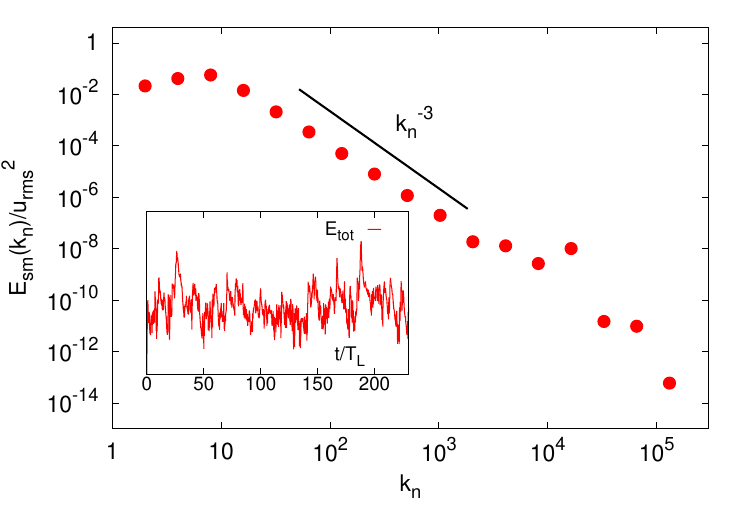}
  \caption{MAIN PANEL: Time-averaged (over the steady state) energy spectrum, 
  normalized by the mean square velocity $u_{\text{rms}}^2 = \overline{\sum_n |u_n|^2}$, from the simulation of the shell model for active turbulence, Eq.~(\ref{eq:sabract1}). INSET: Total energy, $E_{\text{tot}}(t) = \sum_n |u_n|^2$, vs time (in units of the integral scale characteristic time $T_L=(k_{\text{max}} u_{\text{rms}})^{-1}$).}
\label{fig:spectrum-sm}
\end{center}
\end{figure}

\noindent One may wonder whether and to which extent it is possible 
to extend concepts and tools 
developed for inertial turbulence to active turbulence. 
To this end, we introduce a {\it shell model} for active turbulence that 
will generalize and rationalize the observed results.
Shell models (SMs) are deterministic dynamical systems that reduce the complexity of the full (field)
equations, though retaining some of their essential features. Originally introduced as a proxy of the Navier-Stokes
equations, 
they represent a low number of degrees-of-freedom description of hydrodynamic turbulence. 
As such, they offer the possibility to investigate the chaotic dynamics and multiscale correlations
of turbulence with obvious computational advantages ~\cite{GledzerSPD1973,DesnyanskyIANSFAO1974,Frisch,Bohr,BiferaleARFM2003}.
Mathematically, SMs consist of a set of coupled ordinary differential equations, describing the time evolution of complex variables, $\tilde{u}_n(t)$ (with $n=1,2,\dots,N_s$),
that can be thought of as Fourier amplitudes of velocity fluctuations over a length scale with associated wavenumber $k_n$. SMs do not carry any of the {\it geometrical} information contained in the original system, as $\tilde{u}_n$ and $k_n$ are both scalar variables. $k_n$ stands for a radial coordinate in spectral space,
whence the name {\it shell} (correspondingly, the discrete index $n$ is called shell index). Recently, SMs have been extended to  polymeric solutions, to address 
drag reduction and elastic turbulence~\cite{BenziPRE2003,BenziEPL2004,RayEPL2016}\footnote{The system for the velocity variables is coupled, in this case, to a set of equations for analogous
{\it polymer} variables, interpreted as spectral amplitudes of an auxiliary vector field whose dyadic product is the
polymer conformation tensor~\cite{BenziPRE2003}.}.
Hydrodynamic theories describe active suspensions in terms of  the bacteria concentration field, $c(\mathbf{x},t)$, the (incompressible) fluid velocity field, $\mathbf{u}(\mathbf{x},t)$, and an order parameter 
quantifying the degree of local orientation $\mathbf{p}(\mathbf{x},t)$, which represents the average, within a fluid element, of  the particle intrinsic swimming director,  $p_i \propto \langle \hat{e}_i \rangle$,
At odds with concentrated conditions, where the lack of alignment interactions rules 
out the local orientation $\mathbf{p}$ as an appropriate field variable, in semi-dilute suspensions
hydrodynamic interactions can induce locally (and, under certain circumstances, even globally) 
polar order~\cite{EvansPOF2011,AlarconJML2013,YoshinagaPRE2017}.
If we  assume, for simplicity, that the microswimmers spatial distribution remains homogeneous in time ($c(\mathbf{x},t) \approx c_0$, which is indeed confirmed by the DNS; 
consequently we also assume the field $\mathbf{p}$ to be divergence-free), the equations of motion read~\cite{SimhaPRL2002,HatwalnePRL2004,SaintillianPRL2008,SaintillianPOF2008,BaskaranPNAS2009},
\begin{align}\label{eq:eom}
      \rho(\partial_t \mathbf{u} + \mathbf{u} \cdot \nabla) \mathbf{u} &= -\nabla P + \eta \nabla^2 \mathbf{u} + \nabla \cdot \mathbf{\sigma}^{(a)} \nonumber \\ 
\partial_t \mathbf{p} + (\mathbf{u} + w \mathbf{p}) \cdot \nabla \mathbf{p} &= \mathbf{\Omega} \cdot \mathbf{p} 
  - \Gamma \mathbf{p} + D \nabla^2 \mathbf{p};
\end{align}
 $\sigma^{(a)}_{ij} = \zeta p_i p_j $~\cite{HatwalnePRL2004} is the active stress, 
 where $\zeta$ is proportional to 
the swimmers' volume fraction and to the amplitude of the generated stresslet and, therefore, quantifies the level of activity. $\Omega_{ij} = \frac{1}{2}(\partial_i u_j - \partial_j u_i)$ denotes the antisymmetric part of 
the velocity gradient tensor,  and $\eta$ and $D$ are
the fluid dynamic viscosity and the orientation field diffusion coefficient, respectively.
Except for the term $w (\mathbf{p}\cdot \nabla) \mathbf{p}$, 
Eqs.~(\ref{eq:eom}) closely resemble the Oldroyd-B equations for the polymer conformation~\cite{Bird}, in vectorial form~\cite{BenziPRE2003}.\\
Inspired by this formal analogy, we can write a shell model for active fluids, with the inclusion of the self-propulsion,
that will couple the dynamics of $\tilde{u}_n$ with an equation for $\tilde{p}_n$,
the amplitude of orientation magnitude fluctuations at the wavenumber $k_n$.
To this aim, we first need to introduce the following operator~\cite{BenziEPL2004} 
(we will omit hereafter the tilde $\tilde{(\dots)}$, for the sake of lightening the notation)
\begin{align}\label{eq:sabra}
  \Phi^{(\varepsilon)}_n(u,v) &= k_n \left[(1-\varepsilon)u_{n+2} v^{\ast}_{n+1} + (2+\varepsilon)u^{\ast}_{n+1}v_{n+2}\right] + \nonumber \\
                       &= k_{n-1} \left[(2\varepsilon+1)u^{\ast}_{n-1} v_{n+1} - (1-\varepsilon)u_{n+1}v^{\ast}_{n-1}\right] + \nonumber \\ 
                       &= k_{n-2} \left[(2+\varepsilon)u_{n-1} v_{n-2} + (2\varepsilon+1)u_{n-2}v_{n-1}\right] 
\end{align}
which is required, in the shell model, to account for the nonlinear terms in (\ref{eq:eom}). Eq.~(\ref{eq:sabra}) is the generalization to two arguments of the nonlinear coupling
(giving rise to the energy flux in spectral space) of the so called 
``Sabra model''~\cite{LvovPRE1998}. In terms of (\ref{eq:sabra}),
the shell model for the Eqs.~(\ref{eq:eom}) can be written as:
\begin{align}\label{eq:sabract}
  \dot{u}_n  &=  \frac{i}{3} \Phi^{(\varepsilon)}_n(u,u) - \gamma_u(k_n) u_n + f^{(a)}_n \nonumber \\
  \dot{p}_n  &=  \frac{i}{3}\Phi^{(\varepsilon)}_n(u,p) + \frac{i}{3} w \Phi^{(\varepsilon)}_n(p,p) -\frac{i}{3} \Phi^{(\varepsilon)}_n(p,u) \nonumber \\ 
  &-\gamma_p(k_n)p_n - \Gamma p_n.
\end{align}
The set of wavenumbers is taken to be $k_n = k_02^n$. The linear damping terms stem from dissipation ($\gamma_u(k_n)u_n$) and diffusion ($\gamma_p(k_n)p_n$),
and the coefficients read $\gamma_u(k_n) = \nu_{u,p} k_n^2 + \mu_{u,p} k_n^{-4}$, where $\nu_{u,p}$ are the actual viscosity and diffusion coefficient, acting
at small scales (large wavenumbers), whereas $\mu_{u,p}$ are large scale drag coefficients, mimicking friction with the boundaries~\cite{BenziEPL2004,GilbertPRL2002}.
The parameter $\varepsilon$ determines the sign of the flux of generalized ``energy'' across shells, i.e. whether the kinetic energy, $|u_n|^2$, or, alternatively,
the orientation magnitude, $|p_n|^2$, are transferred from large to small scales (as it is, for instance, in actual inertial, three-dimensional, turbulence) or vice versa.
In particular, for $\varepsilon_c < \varepsilon < 0$ downwards transfer is supported (the so called {\it direct cascade}), whereas for 
$\varepsilon < \varepsilon_c = -1 -2^{-2/3}$~\cite{GilbertPRL2002}
upwards transfer (the {\it inverse cascade}) takes place. 
Deciding which is the direction of transfer of fluctuations is not an obvious task; an analog of the Richardson
picture~\cite{Frisch}, that implies a direct cascade in $3d$ inertial turbulence, is, in fact, missing for active turbulence. 
We need, then, to propose an equivalent
phenomenology that goes as follows. First of all, let us notice that, owing to the typically viscous character of collective phenomena in active fluids,
no transfer whatsoever can be ascribed to the non-linear term in the shell model version of the Navier-Stokes equation.
To enforce the zero Reynolds number condition, therefore, we set $\Phi_n^{(\varepsilon)}(u,u)=0 \quad \forall n$ in (\ref{eq:sabract}). The velocity dynamics in active turbulence is, actually,
governed by a scale-matched balance of active forcing and viscous 
dissipation~\cite{CarenzaPRF2020,AlertNP2020}; therefore, to the divergence of the active stress $\sigma^{(a)}$ in (\ref{eq:eom}), a {\it local-in-scale} (or, equivalently, in wavenumber)
force must correspond in the shell model, whence $\nabla \cdot \sigma^{(a)} \rightarrow f^{(a)}_n = i k_n \zeta p_n^2$.
A different mechanism, promoting the emergence of collective motion in the system and hence involving the $P$-equation, should, then, excite the multiple scales in the fluid, i.e. generates the {\it turbulence}.
Such a mechanism can be identified in the flow alignment, whereby the velocity gradients are coupled to the orientational degrees of freedom of the 
microswimmers~\cite{SimhaPRL2002,HernandezOrtizPRL2005,SaintillianPRL2007}. 
We conjecture, then, that the {\it rotation} term, $p\nabla u$, is the one responsible for the upwards transfer and we set $\varepsilon = \varepsilon_b < \varepsilon_c$ in the
operator $\Phi^{(\varepsilon)}_n(p,u)$ in (\ref{eq:sabract}). On the other hand, we expect that the advective, $u\nabla p$, and self-advective, $p \nabla p$, terms have mixing 
properties that should tend to disrupt
spatial coherence of orientation and, therefore, to transfer downwards.
We want to justify empirically these assumptions {\it a posteriori},
benchmarking them against direct numerical simulations. 
Since our numerical method couples a Lagrangian dynamics for microswimmers with a Eulerian description of 
the fluid, though, first we need a 
procedure that maps the particle positions and orientations to the field $\mathbf{P}$. 
The latter is reconstructed with the aid of a Gaussian kernel according to the expression:
\begin{equation}\label{eq:PDelta}
\mathbf{p}(\mathbf{x},t) = \sum_{i=1}^N G(|\mathbf{x}-\mathbf{X}_i(t)|) \hat{\mathbf{e}}_i,
\end{equation}
where $G(\xi) = A e^{-(\xi/R)^2}$ and $A$ is a prefactor such that
$\int G(\xi) d^3 \xi = 1$.  
At this point, by projecting the second of Eqs.~(\ref{eq:eom}) 
onto the Fourier mode $\mathbf{k}$, multiplying both sides by $\tilde{\mathbf{p}}^{\ast}$
(the complex conjugate of the Fourier transform of $\mathbf{p}$), averaging
over shells of radius $k$ and summing with the complex conjugate equation, we get:
\begin{equation}\label{eq:spectralP}
  (\partial_t + \Gamma + D k^2)\mathcal{P} =
  \mathcal{F}_{\text{adv}} + \mathcal{F}_{\text{self}} + \mathcal{F}_{\text{rot}},
\end{equation}
where $\mathcal{P}(k,t) = \langle |\tilde{\mathbf{p}}|^2 \rangle$. The terms on the right hand 
side read as follows:
\begin{align}\label{eq:fluxes}
\mathcal{F}_{\text{adv}}(k,t) &= -\langle \left(\tilde{\mathbf{p}}^{\ast}(\mathbf{k},t)\cdot \mathbf{J}_{\text{up}}(\mathbf{k},t) \right)\rangle + \mbox{c.c.} \\ \nonumber
\mathcal{F}_{\text{self}}(k,t) &= -\langle \left(\tilde{\mathbf{p}}^{\ast}(\mathbf{k},t)\cdot \mathbf{J}_{\text{pp}}(\mathbf{k},t) \right)\rangle + \mbox{c.c.} \\ \nonumber
\mathcal{F}_{\text{rot}}(k,t) &= \langle \left(\tilde{\mathbf{p}}^{\ast}(\mathbf{k},t)\cdot \mathbf{\mathcal{R}}(\mathbf{k},t) \right)\rangle + \mbox{c.c.}
\end{align}
where $\mathbf{J}_{\text{up}}(\mathbf{k},t)$, $\mathbf{J}_{\text{pp}}(\mathbf{k},t)$ and 
$\mathbf{\mathcal{R}}(\mathbf{k},t)$ are the Fourier transforms of the non-linear terms of 
the orientation field equation, namely $\mathbf{u} \cdot \nabla \mathbf{p}$,
$w\mathbf{p}\cdot \mathbf{p}$ and 
$\mathbf{\Omega} \cdot \mathbf{p}$, respectively, and "$\mbox{c.c.}$" stands for the complex conjugate 
terms. 
Eqs.~(\ref{eq:fluxes}) are fluxes across $k$-shells in spectral space and whether a direct or inverse {\it cascade of magnitude of orientation}, depending on their sign, takes place.
\begin{figure}
\begin{center}
  \advance\leftskip-0.55cm
  \includegraphics[scale=0.8]{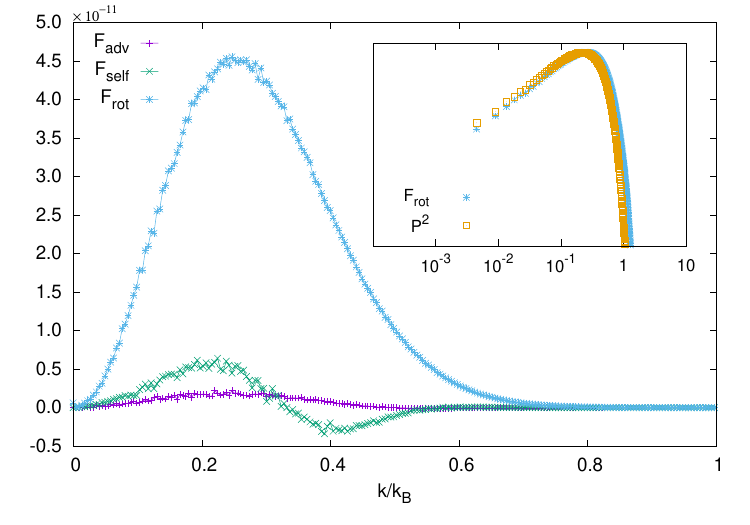}
  \caption{Spectral fluxes, Eq.~(\ref{eq:fluxes}), measured from the numerical simulations (time-averaged over the statistically stationary state). 
  In the inset we check the expected proportionality between $\mathcal{F}_{\text{rot}}$ and $\mathcal{P}$, thus 
  validating Eq.~(\ref{eq:sm_balance}) in the direct numerical simulations ($\mathcal{P}$ is scaled by a factor $\sim 2.6 \times 10^{-4}$).}
\label{fig:fluxes} 
\end{center}
\end{figure}
We measured them in the numerical simulations and the results, averaged in time over the statistically stationary state are plotted in Fig.~\ref{fig:fluxes}. 
We see that, while $\mathcal{F}_{\text{adv}}$ and $\mathcal{F}_{\text{self}}$ (top and middle panels) are negative at intermediate and large wavenumbers, signalling that they transfer orientation fluctuations towards small scales (i.e. they tend to disrupt coherence), the rotational spectral flux $\mathcal{F}_{\text{rot}}$ (bottom panel) is positive (and in magnitude much larger than the previous two), therefore confirming our phenomenological conjecture that it is the term responsible for the upward {\it cascade} that eventually pumps energy into the large scales.
We set, then, $\varepsilon = \varepsilon_f > \varepsilon_c$ in the operators
$\Phi^{(\varepsilon)}_n(u,p)$ and $\Phi^{(\varepsilon)}_n(p,p)$ in (\ref{eq:sabract}).
Eventually, we arrive at the following structure for the sought {\it SabrActive} shell model:
\begin{align}\label{eq:sabract1}
  \dot{u}_n & =  -\gamma_u(k_n) u_n + i  \zeta k_n p_n^2 \nonumber \\
  \dot{p}_n & =  \frac{i}{3}\Phi^{(\varepsilon_f)}_n(u,p) + \frac{i}{3} w \Phi^{(\varepsilon_f)}_n(p,p) -\frac{i}{3} \Phi^{(\varepsilon_b)}_n(p,u) \nonumber \\
            &  -\gamma_p(k_n)p_n - \Gamma p_n + \delta_{n,n_B}|p_n|^{-1}p_n p_B.
\end{align}
The {\it forcing} term $\delta_{n,n_B}|p_n|^{-1}p_n p_B$ has been introduced in the $p$-equation  
to account for the fact that at the smallest scale (ideally that of a bacterium) the
orientation is fixed by the intrinsic microorganism swimming direction.\\ 
We integrated the system (\ref{eq:sabract1}) with $N_s = 20$ shells by means of a fourth-order Runge-Kutta scheme for $T=6 \times 10^{11}$ 
time steps (with integration step $\Delta t = 10^{-4}$), corresponding
to $\approx 200 T_L$, $T_L=(k_{\text{max}} u_{\text{rms}})^{-1}$ being the large scale characteristic time, where $u_{\text{rms}} = \left(\sum_n |u_n|^2\right)^{1/2}$ and $k_{\text{max}}$ is the location of spectrum maximum (i.e. the 
wavenumber of the energy-containing scales)~\cite{PisarenkoPOF1993}. 
The following numerical values are used for the parameters: $k_0=2^{-4}$, $\varepsilon_f=-0.4$, $\varepsilon_b=-1.8$, $\nu_u=10^{-6}$, $\mu_u=10^{-10}$, $\zeta=1.25 \times 10^{-2}$, $w=1.25 \times 10^{-2}$,
$\nu_p = 8.5 \times 10^{-13}$, $\mu_p=10^{-10}$, $\Gamma = 10^{-6}$, $p_B = 5 \times 10^{-11} (1 + i)$, $n_B=N_s-1$.\\
Fig.~\ref{fig:spectrum-sm} displays the energy spectrum, $E_{\text{sm}}(k_n) = \langle \frac{|u_n|^2}{k_n} \rangle$, for the shell model,
where the average, $\langle (\cdots) \rangle$ is meant taken over time, in the statistically stationary state $t \gtrapprox 10 T_L$ 
(see inset of Fig.~\ref{fig:spectrum-sm} displaying the total energy, 
$E_{\text{tot}}(t) = \sum_n |u_n|^2$, as a function of time).
The spectrum decays over a quite wide range of wavenumbers as $E_{\text{sm}}(k_n) \sim k_n^{-3}$ 
(solid line in Fig.~\ref{fig:spectrum-sm}), in agreement with
the DNS. 
Moreover, the shell model allows to provide a theoretical explanation of such power law decay. 
To this aim, let us obtain the evolution of the orientation magnitude, $|p_n|^2$, by multiplying
the second of Eqs.~(\ref{eq:sabract1}) by $p_n^{\ast}$ and its
complex conjugate by $p_n$ and summing the two:
\begin{equation}
  \partial_t |p_n|^2 \approx p_n^{\ast}\Phi^{(\varepsilon_b)}_n(p,u) + p_n(\Phi^{(\varepsilon_b)}_n(p,u))^{\ast} - \Gamma |p_n|^2 -\gamma(k_n)|p_n|^2.
\end{equation}  
Here the advective and self-advecting terms have been omitted, since they are much smaller 
than the rotation, as observed in the (see Fig.~\ref{fig:fluxes}).
Assuming statistical stationarity and focusing on intermediate $k_n$, where diffusive terms 
can be neglected, we see that 
\begin{equation}\label{eq:sm_balance}
k u p^2 \sim \Gamma p^2.
\end{equation}
From Eq.~(\ref{eq:sm_balance}) we get $u \sim k^{-1}$ and, then, the energy spectrum, $k_n^{-1}|u_n|^2$, should indeed behave as 
\begin{equation}\label{eq:sf2-theo}
  E_{\text{sm}}(k_n) \sim k_n^{-3}.
\end{equation}

\section{Conclusions}
\noindent We have presented a computational and theoretical study aimed at revealing the presence of 
active turbulence in pusher suspensions in conditions of semi-dilution, 
namely far from close packing but above the onset of collective motion.
We reported that the Eulerian solvent velocity is 
L\'evy-distributed,
similarly to previous experimental and theoretical results~\cite{RushkinPRL2010,ZaidJRSI2011}.
We showed that the energy spectrum develops decays with the power-law 
$k^{-3}$ over a range of intermediate wavenumbers. It was posited that, given the lack of a
Richardson-Kolmogorov energy cascade as in classical turbulence, the excitation of motion at large 
scales should be ascribed to the coupling between fluid velocity gradients and 
microswimmer orientation, i.e. to the flow alignment mechanism. 
Based on this picture and on phenomenological arguments, we developed a reduce order dynamical 
deterministic model (or {\it shell model}) of active turbulence, dubbed {\it SabrActive model}. 
Numerical simulations and theoretical analysis of the model confirmed the $k^{-3}$ scaling of the spectrum.\\ 
The introduction of this new model pushes forward the reach of quantitative tests of how actually "turbulent" is active turbulence, allowing to measure, e.g., Lyapunov exponents, higher order structure functions,
multiscale statistics, etc, on "physically" much longer runs. A flavour of this capability can be grasped in the observation of intermittency in the PDFs of shell model velocity variables.
The insight provided  will motivate further analysis of  the dynamical and statistical properties of the model and the sensitivity of the system response to changes in the control parameters
(for instance, the onset of active turbulence at changing the "activity parameter"), as well as exploring 
a wider region of the volume-fraction/squirming-parameters space in DNS.\\
\section*{Acknowledgements}
\noindent I.P. acknowledges support from Ministerio de Ciencia, Innovación y
Universidades MCIU/AEI/FEDER for financial support under
grant agreement PID2021-126570NB-100 AEI/FEDER-EU, from
Generalitat de Catalunya  under Program Icrea Acad\`emia and project 2021SGR-673.
This work was possible thanks to the access to the MareNostrum Supercomputer at Barcelona
Supercomputing Center (BSC) and also through the Partnership
for Advanced Computing in Europe (PRACE).

\bibliography{gps_prfluids}

\begin{thebibliography}{73}%
\makeatletter
\providecommand \@ifxundefined [1]{%
 \@ifx{#1\undefined}
}%
\providecommand \@ifnum [1]{%
 \ifnum #1\expandafter \@firstoftwo
 \else \expandafter \@secondoftwo
 \fi
}%
\providecommand \@ifx [1]{%
 \ifx #1\expandafter \@firstoftwo
 \else \expandafter \@secondoftwo
 \fi
}%
\providecommand \natexlab [1]{#1}%
\providecommand \enquote  [1]{``#1''}%
\providecommand \bibnamefont  [1]{#1}%
\providecommand \bibfnamefont [1]{#1}%
\providecommand \citenamefont [1]{#1}%
\providecommand \href@noop [0]{\@secondoftwo}%
\providecommand \href [0]{\begingroup \@sanitize@url \@href}%
\providecommand \@href[1]{\@@startlink{#1}\@@href}%
\providecommand \@@href[1]{\endgroup#1\@@endlink}%
\providecommand \@sanitize@url [0]{\catcode `\\12\catcode `\$12\catcode
  `\&12\catcode `\#12\catcode `\^12\catcode `\_12\catcode `\%12\relax}%
\providecommand \@@startlink[1]{}%
\providecommand \@@endlink[0]{}%
\providecommand \url  [0]{\begingroup\@sanitize@url \@url }%
\providecommand \@url [1]{\endgroup\@href {#1}{\urlprefix }}%
\providecommand \urlprefix  [0]{URL }%
\providecommand \Eprint [0]{\href }%
\providecommand \doibase [0]{http://dx.doi.org/}%
\providecommand \selectlanguage [0]{\@gobble}%
\providecommand \bibinfo  [0]{\@secondoftwo}%
\providecommand \bibfield  [0]{\@secondoftwo}%
\providecommand \translation [1]{[#1]}%
\providecommand \BibitemOpen [0]{}%
\providecommand \bibitemStop [0]{}%
\providecommand \bibitemNoStop [0]{.\EOS\space}%
\providecommand \EOS [0]{\spacefactor3000\relax}%
\providecommand \BibitemShut  [1]{\csname bibitem#1\endcsname}%
\let\auto@bib@innerbib\@empty
\bibitem [{\citenamefont {Vicsek}\ and\ \citenamefont
  {Zafeiris}(2012)}]{VicsekPR2012}%
  \BibitemOpen
  \bibfield  {author} {\bibinfo {author} {\bibfnamefont {T}~\bibnamefont
  {Vicsek}}\ and\ \bibinfo {author} {\bibfnamefont {A.}~\bibnamefont
  {Zafeiris}},\ }\bibfield  {title} {\enquote {\bibinfo {title} {Collective
  motion},}\ }\href@noop {} {\bibfield  {journal} {\bibinfo  {journal} {Phys.
  Rep.}\ }\textbf {\bibinfo {volume} {517}},\ \bibinfo {pages} {71--149}
  (\bibinfo {year} {2012})}\BibitemShut {NoStop}%
\bibitem [{\citenamefont {Kearns}(2010)}]{KearnsNature2010}%
  \BibitemOpen
  \bibfield  {author} {\bibinfo {author} {\bibfnamefont {D.B.}\ \bibnamefont
  {Kearns}},\ }\bibfield  {title} {\enquote {\bibinfo {title} {A field guide to
  bacterial swarming motility},}\ }\href@noop {} {\bibfield  {journal}
  {\bibinfo  {journal} {Nat. Rev. Microbiol.}\ }\textbf {\bibinfo {volume}
  {8}},\ \bibinfo {pages} {634--644} (\bibinfo {year} {2010})}\BibitemShut
  {NoStop}%
\bibitem [{\citenamefont {Darnton}\ \emph {et~al.}(2010)\citenamefont
  {Darnton}, \citenamefont {Turner}, \citenamefont {Rojevsky},\ and\
  \citenamefont {Berg}}]{DarntonBJ2010}%
  \BibitemOpen
  \bibfield  {author} {\bibinfo {author} {\bibfnamefont {N.C.}\ \bibnamefont
  {Darnton}}, \bibinfo {author} {\bibfnamefont {L.}~\bibnamefont {Turner}},
  \bibinfo {author} {\bibfnamefont {S.}~\bibnamefont {Rojevsky}}, \ and\
  \bibinfo {author} {\bibfnamefont {H.C.}\ \bibnamefont {Berg}},\ }\bibfield
  {title} {\enquote {\bibinfo {title} {Dynamics of bacterial swarming},}\
  }\href@noop {} {\bibfield  {journal} {\bibinfo  {journal} {Biophys. J.}\
  }\textbf {\bibinfo {volume} {98}},\ \bibinfo {pages} {2082--2090} (\bibinfo
  {year} {2010})}\BibitemShut {NoStop}%
\bibitem [{\citenamefont {Koch}\ and\ \citenamefont
  {Subramanian}(2011)}]{KochARFM2011}%
  \BibitemOpen
  \bibfield  {author} {\bibinfo {author} {\bibfnamefont {D.L.}\ \bibnamefont
  {Koch}}\ and\ \bibinfo {author} {\bibfnamefont {G.}~\bibnamefont
  {Subramanian}},\ }\bibfield  {title} {\enquote {\bibinfo {title} {Collective
  hydrodynamics of swimming microorganisms: living fluids},}\ }\href@noop {}
  {\bibfield  {journal} {\bibinfo  {journal} {Annu. Rev. Fluid Mech.}\ }\textbf
  {\bibinfo {volume} {43}},\ \bibinfo {pages} {637--659} (\bibinfo {year}
  {2011})}\BibitemShut {NoStop}%
\bibitem [{\citenamefont {Ballerini}\ \emph {et~al.}(2008)\citenamefont
  {Ballerini}, \citenamefont {Cabibbo}, \citenamefont {Candelier},
  \citenamefont {Cavagna}, \citenamefont {Cisbani}, \citenamefont {Giardina},
  \citenamefont {Lecomte}, \citenamefont {Orlandi}, \citenamefont {Parisi},
  \citenamefont {Procaccini}, \citenamefont {Viale},\ and\ \citenamefont
  {Zdravkovic}}]{BalleriniPNAS2008}%
  \BibitemOpen
  \bibfield  {author} {\bibinfo {author} {\bibfnamefont {M.}~\bibnamefont
  {Ballerini}}, \bibinfo {author} {\bibfnamefont {N.}~\bibnamefont {Cabibbo}},
  \bibinfo {author} {\bibfnamefont {R.}~\bibnamefont {Candelier}}, \bibinfo
  {author} {\bibfnamefont {A.}~\bibnamefont {Cavagna}}, \bibinfo {author}
  {\bibfnamefont {E.}~\bibnamefont {Cisbani}}, \bibinfo {author} {\bibfnamefont
  {I.}~\bibnamefont {Giardina}}, \bibinfo {author} {\bibfnamefont
  {V.}~\bibnamefont {Lecomte}}, \bibinfo {author} {\bibfnamefont
  {A.}~\bibnamefont {Orlandi}}, \bibinfo {author} {\bibfnamefont
  {G.}~\bibnamefont {Parisi}}, \bibinfo {author} {\bibfnamefont
  {A.}~\bibnamefont {Procaccini}}, \bibinfo {author} {\bibfnamefont
  {M.}~\bibnamefont {Viale}}, \ and\ \bibinfo {author} {\bibfnamefont
  {V.}~\bibnamefont {Zdravkovic}},\ }\bibfield  {title} {\enquote {\bibinfo
  {title} {Interaction ruling animal collective behavior depends on topological
  rather than metric distance: Evidence from a field study},}\ }\href@noop {}
  {\bibfield  {journal} {\bibinfo  {journal} {Proc. Natl. Acad. Sci. USA}\
  }\textbf {\bibinfo {volume} {105}},\ \bibinfo {pages} {1232--1237} (\bibinfo
  {year} {2008})}\BibitemShut {NoStop}%
\bibitem [{\citenamefont {Herbert-Read}\ \emph {et~al.}(2011)\citenamefont
  {Herbert-Read}, \citenamefont {Perna}, \citenamefont {Mann}, \citenamefont
  {Schaerf}, \citenamefont {Sumpter},\ and\ \citenamefont
  {Ward}}]{HerbertPNAS2011}%
  \BibitemOpen
  \bibfield  {author} {\bibinfo {author} {\bibfnamefont {J.E.}\ \bibnamefont
  {Herbert-Read}}, \bibinfo {author} {\bibfnamefont {A.}~\bibnamefont {Perna}},
  \bibinfo {author} {\bibfnamefont {R.P.}\ \bibnamefont {Mann}}, \bibinfo
  {author} {\bibfnamefont {T.M.}\ \bibnamefont {Schaerf}}, \bibinfo {author}
  {\bibfnamefont {D.J.T.}\ \bibnamefont {Sumpter}}, \ and\ \bibinfo {author}
  {\bibfnamefont {A.J.W.}\ \bibnamefont {Ward}},\ }\bibfield  {title} {\enquote
  {\bibinfo {title} {Inferring the rules of interaction of shoaling fish},}\
  }\href@noop {} {\bibfield  {journal} {\bibinfo  {journal} {Proc. Natl. Acad.
  Sci. USA}\ }\textbf {\bibinfo {volume} {108}},\ \bibinfo {pages}
  {18726--18731} (\bibinfo {year} {2011})}\BibitemShut {NoStop}%
\bibitem [{\citenamefont {Dombrowski}\ \emph {et~al.}(2004)\citenamefont
  {Dombrowski}, \citenamefont {Cisneros}, \citenamefont {Chatkaew},
  \citenamefont {Goldstein},\ and\ \citenamefont
  {Kessler}}]{DombrowskiPRL2004}%
  \BibitemOpen
  \bibfield  {author} {\bibinfo {author} {\bibfnamefont {C.}~\bibnamefont
  {Dombrowski}}, \bibinfo {author} {\bibfnamefont {L.}~\bibnamefont
  {Cisneros}}, \bibinfo {author} {\bibfnamefont {S.}~\bibnamefont {Chatkaew}},
  \bibinfo {author} {\bibfnamefont {R.E.}\ \bibnamefont {Goldstein}}, \ and\
  \bibinfo {author} {\bibfnamefont {J.O.}\ \bibnamefont {Kessler}},\ }\bibfield
   {title} {\enquote {\bibinfo {title} {Self-concentration and large-scale
  coherence in bacterial dynamics},}\ }\href@noop {} {\bibfield  {journal}
  {\bibinfo  {journal} {Phys. Rev. Lett.}\ }\textbf {\bibinfo {volume} {93}},\
  \bibinfo {pages} {098103} (\bibinfo {year} {2004})}\BibitemShut {NoStop}%
\bibitem [{\citenamefont {Wolgemuth}(2008)}]{WolgemuthBJ2008}%
  \BibitemOpen
  \bibfield  {author} {\bibinfo {author} {\bibfnamefont {C.W.}\ \bibnamefont
  {Wolgemuth}},\ }\bibfield  {title} {\enquote {\bibinfo {title} {Collective
  swimming and the dynamics of bacterial turbulence},}\ }\href@noop {}
  {\bibfield  {journal} {\bibinfo  {journal} {Biophys. J.}\ }\textbf {\bibinfo
  {volume} {95}},\ \bibinfo {pages} {1564--1574} (\bibinfo {year}
  {2008})}\BibitemShut {NoStop}%
\bibitem [{\citenamefont {Pedley}\ and\ \citenamefont
  {Kessler}(1992)}]{PedleyARFM1992}%
  \BibitemOpen
  \bibfield  {author} {\bibinfo {author} {\bibfnamefont {T.J.}\ \bibnamefont
  {Pedley}}\ and\ \bibinfo {author} {\bibfnamefont {J.O.}\ \bibnamefont
  {Kessler}},\ }\bibfield  {title} {\enquote {\bibinfo {title} {Hydrodynamic
  phenomena in suspensions of swimming microorganisms},}\ }\href@noop {}
  {\bibfield  {journal} {\bibinfo  {journal} {Annu. Rev. Fluid Mech.}\ }\textbf
  {\bibinfo {volume} {23}},\ \bibinfo {pages} {313--358} (\bibinfo {year}
  {1992})}\BibitemShut {NoStop}%
\bibitem [{\citenamefont {Alert}\ and\ \citenamefont
  {Casademunt}(2020)}]{AlertARCMP2022}%
  \BibitemOpen
  \bibfield  {author} {\bibinfo {author} {\bibfnamefont {R.}~\bibnamefont
  {Alert}}\ and\ \bibinfo {author} {\bibfnamefont {J.-F.}\ \bibnamefont
  {Casademunt}, \bibfnamefont {J.~Joanny}},\ }\bibfield  {title} {\enquote
  {\bibinfo {title} {Active turbulence},}\ }\href@noop {} {\bibfield  {journal}
  {\bibinfo  {journal} {Annu. Rev. Condens. Matter Phys.}\ }\textbf {\bibinfo
  {volume} {13}},\ \bibinfo {pages} {143--170} (\bibinfo {year}
  {2020})}\BibitemShut {NoStop}%
\bibitem [{\citenamefont {Vallis}(1999)}]{Vallis}%
  \BibitemOpen
  \bibfield  {author} {\bibinfo {author} {\bibfnamefont {G.K.}\ \bibnamefont
  {Vallis}},\ }\href@noop {} {\emph {\bibinfo {title} {Lecture Notes on
  "Geostrophys turbulence: the macroturbulence of the atmosphere and ocean"}}}\
  (\bibinfo {year} {1999})\BibitemShut {NoStop}%
\bibitem [{\citenamefont {Ishikawa}\ \emph {et~al.}(2011)\citenamefont
  {Ishikawa}, \citenamefont {Yoshida}, \citenamefont {Uedo}, \citenamefont
  {Wiedemann}, \citenamefont {Imai},\ and\ \citenamefont
  {Yamaguchi}}]{IshikawaPRL2011}%
  \BibitemOpen
  \bibfield  {author} {\bibinfo {author} {\bibfnamefont {T.}~\bibnamefont
  {Ishikawa}}, \bibinfo {author} {\bibfnamefont {N.}~\bibnamefont {Yoshida}},
  \bibinfo {author} {\bibfnamefont {H.}~\bibnamefont {Uedo}}, \bibinfo {author}
  {\bibfnamefont {M.}~\bibnamefont {Wiedemann}}, \bibinfo {author}
  {\bibfnamefont {Y.}~\bibnamefont {Imai}}, \ and\ \bibinfo {author}
  {\bibfnamefont {T.}~\bibnamefont {Yamaguchi}},\ }\bibfield  {title} {\enquote
  {\bibinfo {title} {Energy transport in a concentrated suspension of
  bacteria},}\ }\href@noop {} {\bibfield  {journal} {\bibinfo  {journal} {Phys.
  Rev. Lett.}\ }\textbf {\bibinfo {volume} {107}},\ \bibinfo {pages} {028102}
  (\bibinfo {year} {2011})}\BibitemShut {NoStop}%
\bibitem [{\citenamefont {Wensink}\ \emph {et~al.}(2012)\citenamefont
  {Wensink}, \citenamefont {Dunkel}, \citenamefont {Heidenreich}, \citenamefont
  {Drescher}, \citenamefont {Goldstein}, \citenamefont {L\"owen},\ and\
  \citenamefont {Yeomans}}]{WensinkPNAS2012}%
  \BibitemOpen
  \bibfield  {author} {\bibinfo {author} {\bibfnamefont {H.H.}\ \bibnamefont
  {Wensink}}, \bibinfo {author} {\bibfnamefont {J.}~\bibnamefont {Dunkel}},
  \bibinfo {author} {\bibfnamefont {S.}~\bibnamefont {Heidenreich}}, \bibinfo
  {author} {\bibfnamefont {K.}~\bibnamefont {Drescher}}, \bibinfo {author}
  {\bibfnamefont {R.E.}\ \bibnamefont {Goldstein}}, \bibinfo {author}
  {\bibfnamefont {H.}~\bibnamefont {L\"owen}}, \ and\ \bibinfo {author}
  {\bibfnamefont {J.M.}\ \bibnamefont {Yeomans}},\ }\bibfield  {title}
  {\enquote {\bibinfo {title} {Meso-scale turbulence in living fluids},}\
  }\href@noop {} {\bibfield  {journal} {\bibinfo  {journal} {Proc. Natl. Acad.
  Sci. USA}\ }\textbf {\bibinfo {volume} {109}},\ \bibinfo {pages}
  {14308--14313} (\bibinfo {year} {2012})}\BibitemShut {NoStop}%
\bibitem [{\citenamefont {Kokot}\ \emph {et~al.}(2015)\citenamefont {Kokot},
  \citenamefont {Das}, \citenamefont {Winkler}, \citenamefont {Gomppper},
  \citenamefont {Aranson},\ and\ \citenamefont {Snezhko}}]{KokotPNAS2017}%
  \BibitemOpen
  \bibfield  {author} {\bibinfo {author} {\bibfnamefont {G.}~\bibnamefont
  {Kokot}}, \bibinfo {author} {\bibfnamefont {S.}~\bibnamefont {Das}}, \bibinfo
  {author} {\bibfnamefont {R.G.}\ \bibnamefont {Winkler}}, \bibinfo {author}
  {\bibfnamefont {G.}~\bibnamefont {Gomppper}}, \bibinfo {author}
  {\bibfnamefont {I.S.}\ \bibnamefont {Aranson}}, \ and\ \bibinfo {author}
  {\bibfnamefont {A.}~\bibnamefont {Snezhko}},\ }\bibfield  {title} {\enquote
  {\bibinfo {title} {Active turbulence in a gas of self--assembled spinners},}\
  }\href@noop {} {\bibfield  {journal} {\bibinfo  {journal} {Proc. Natl. Acad.
  Sci. USA}\ }\textbf {\bibinfo {volume} {112}},\ \bibinfo {pages}
  {15048--15053} (\bibinfo {year} {2015})}\BibitemShut {NoStop}%
\bibitem [{\citenamefont {Doostmohammadi}\ \emph {et~al.}(2017)\citenamefont
  {Doostmohammadi}, \citenamefont {Shendruk}, \citenamefont {Thijssen},\ and\
  \citenamefont {Yeomans}}]{DoostmohammadiNATCOMM2017}%
  \BibitemOpen
  \bibfield  {author} {\bibinfo {author} {\bibfnamefont {A.}~\bibnamefont
  {Doostmohammadi}}, \bibinfo {author} {\bibfnamefont {T.N.}\ \bibnamefont
  {Shendruk}}, \bibinfo {author} {\bibfnamefont {K.}~\bibnamefont {Thijssen}},
  \ and\ \bibinfo {author} {\bibfnamefont {J.M.}\ \bibnamefont {Yeomans}},\
  }\bibfield  {title} {\enquote {\bibinfo {title} {Onset of meso-scale
  turbulence in active nematics},}\ }\href@noop {} {\bibfield  {journal}
  {\bibinfo  {journal} {Nat. Commun.}\ }\textbf {\bibinfo {volume} {8}},\
  \bibinfo {pages} {15326} (\bibinfo {year} {2017})}\BibitemShut {NoStop}%
\bibitem [{\citenamefont {Mart{\'\i}nez-Prat}\ \emph
  {et~al.}(2021)\citenamefont {Mart{\'\i}nez-Prat}, \citenamefont {Alert},
  \citenamefont {Meng}, \citenamefont {Ign{\'e}s-Mullol}, \citenamefont
  {Joanny}, \citenamefont {Casademunt}, \citenamefont {Golestanian},\ and\
  \citenamefont {Sagu{\'e}s}}]{sagues2021}%
  \BibitemOpen
  \bibfield  {author} {\bibinfo {author} {\bibfnamefont {B.}~\bibnamefont
  {Mart{\'\i}nez-Prat}}, \bibinfo {author} {\bibfnamefont {R.}~\bibnamefont
  {Alert}}, \bibinfo {author} {\bibfnamefont {F.}~\bibnamefont {Meng}},
  \bibinfo {author} {\bibfnamefont {J.}~\bibnamefont {Ign{\'e}s-Mullol}},
  \bibinfo {author} {\bibfnamefont {J.-F.}\ \bibnamefont {Joanny}}, \bibinfo
  {author} {\bibfnamefont {J.}~\bibnamefont {Casademunt}}, \bibinfo {author}
  {\bibfnamefont {R.}~\bibnamefont {Golestanian}}, \ and\ \bibinfo {author}
  {\bibfnamefont {F.}~\bibnamefont {Sagu{\'e}s}},\ }\bibfield  {title}
  {\enquote {\bibinfo {title} {Scaling regimes of active turbulence with
  external dissipation},}\ }\href@noop {} {\bibfield  {journal} {\bibinfo
  {journal} {Phys. Rev. X}\ }\textbf {\bibinfo {volume} {11}},\ \bibinfo
  {pages} {031065} (\bibinfo {year} {2021})}\BibitemShut {NoStop}%
\bibitem [{\citenamefont {Bourgoin}\ \emph {et~al.}(2020)\citenamefont
  {Bourgoin}, \citenamefont {Kervil}, \citenamefont {Cottin-Bizonne},
  \citenamefont {Raynal}, \citenamefont {Volk},\ and\ \citenamefont
  {Ybert}}]{BourgoinPRX2020}%
  \BibitemOpen
  \bibfield  {author} {\bibinfo {author} {\bibfnamefont {M.}~\bibnamefont
  {Bourgoin}}, \bibinfo {author} {\bibfnamefont {R.}~\bibnamefont {Kervil}},
  \bibinfo {author} {\bibfnamefont {C.}~\bibnamefont {Cottin-Bizonne}},
  \bibinfo {author} {\bibfnamefont {R.}~\bibnamefont {Raynal}}, \bibinfo
  {author} {\bibfnamefont {R.}~\bibnamefont {Volk}}, \ and\ \bibinfo {author}
  {\bibfnamefont {C.}~\bibnamefont {Ybert}},\ }\bibfield  {title} {\enquote
  {\bibinfo {title} {Kolmogorovian active turbulence of a sparse assembly of
  interacting marangoni surfers},}\ }\href@noop {} {\bibfield  {journal}
  {\bibinfo  {journal} {Phys. Rev. X}\ }\textbf {\bibinfo {volume} {10}},\
  \bibinfo {pages} {021065} (\bibinfo {year} {2020})}\BibitemShut {NoStop}%
\bibitem [{\citenamefont {Bratanov}\ \emph {et~al.}(2015)\citenamefont
  {Bratanov}, \citenamefont {Jenko},\ and\ \citenamefont
  {Frey}}]{BratanovPNAS2015}%
  \BibitemOpen
  \bibfield  {author} {\bibinfo {author} {\bibfnamefont {V.}~\bibnamefont
  {Bratanov}}, \bibinfo {author} {\bibfnamefont {F.}~\bibnamefont {Jenko}}, \
  and\ \bibinfo {author} {\bibfnamefont {E.}~\bibnamefont {Frey}},\ }\bibfield
  {title} {\enquote {\bibinfo {title} {New class of turbulence in active
  fluids},}\ }\href@noop {} {\bibfield  {journal} {\bibinfo  {journal} {Proc.
  Natl. Acad. Sci. USA}\ }\textbf {\bibinfo {volume} {112}},\ \bibinfo {pages}
  {15048--15053} (\bibinfo {year} {2015})}\BibitemShut {NoStop}%
\bibitem [{\citenamefont {Frisch}(1995)}]{Frisch}%
  \BibitemOpen
  \bibfield  {author} {\bibinfo {author} {\bibfnamefont {U.}~\bibnamefont
  {Frisch}},\ }\href@noop {} {\emph {\bibinfo {title} {Turbulence}}}\ (\bibinfo
   {publisher} {Cambridge University Press},\ \bibinfo {year}
  {1995})\BibitemShut {NoStop}%
\bibitem [{\citenamefont {S{\l}omka}\ and\ \citenamefont
  {Dunkel}(2017)}]{SlomkaPNAS2017}%
  \BibitemOpen
  \bibfield  {author} {\bibinfo {author} {\bibfnamefont {L.}~\bibnamefont
  {S{\l}omka}}\ and\ \bibinfo {author} {\bibfnamefont {J.}~\bibnamefont
  {Dunkel}},\ }\bibfield  {title} {\enquote {\bibinfo {title} {Spontaneous
  mirror-symmetry breaking induces inverse energy cascade in 3d active
  fluids},}\ }\href@noop {} {\bibfield  {journal} {\bibinfo  {journal} {Proc.
  Natl. Acad. Sci. USA}\ }\textbf {\bibinfo {volume} {114}},\ \bibinfo {pages}
  {2119--2124} (\bibinfo {year} {2017})}\BibitemShut {NoStop}%
\bibitem [{\citenamefont {Linkmann}\ \emph {et~al.}(2019)\citenamefont
  {Linkmann}, \citenamefont {Boffetta}, \citenamefont {Marchetti},\ and\
  \citenamefont {Eckhardt}}]{LinkmannPRL2019}%
  \BibitemOpen
  \bibfield  {author} {\bibinfo {author} {\bibfnamefont {M.}~\bibnamefont
  {Linkmann}}, \bibinfo {author} {\bibfnamefont {G.}~\bibnamefont {Boffetta}},
  \bibinfo {author} {\bibfnamefont {M.C.}\ \bibnamefont {Marchetti}}, \ and\
  \bibinfo {author} {\bibfnamefont {B.}~\bibnamefont {Eckhardt}},\ }\bibfield
  {title} {\enquote {\bibinfo {title} {Phase transition to large scale coherent
  structures in two-dimensional active matter turbulence},}\ }\href@noop {}
  {\bibfield  {journal} {\bibinfo  {journal} {Phys. Rev. Lett.}\ }\textbf
  {\bibinfo {volume} {122}},\ \bibinfo {pages} {214503} (\bibinfo {year}
  {2019})}\BibitemShut {NoStop}%
\bibitem [{\citenamefont {Carenza}\ \emph
  {et~al.}(2020{\natexlab{a}})\citenamefont {Carenza}, \citenamefont
  {Biferale},\ and\ \citenamefont {Gonnella}}]{CarenzaPRF2020}%
  \BibitemOpen
  \bibfield  {author} {\bibinfo {author} {\bibfnamefont {L.N.}\ \bibnamefont
  {Carenza}}, \bibinfo {author} {\bibfnamefont {L.}~\bibnamefont {Biferale}}, \
  and\ \bibinfo {author} {\bibfnamefont {G.}~\bibnamefont {Gonnella}},\
  }\bibfield  {title} {\enquote {\bibinfo {title} {Multiscale control of active
  emulsion dynamics},}\ }\href@noop {} {\bibfield  {journal} {\bibinfo
  {journal} {Phys. Rev. Fluid}\ }\textbf {\bibinfo {volume} {5}},\ \bibinfo
  {pages} {011302(R)} (\bibinfo {year} {2020}{\natexlab{a}})}\BibitemShut
  {NoStop}%
\bibitem [{\citenamefont {Carenza}\ \emph
  {et~al.}(2020{\natexlab{b}})\citenamefont {Carenza}, \citenamefont
  {Biferale},\ and\ \citenamefont {Gonnella}}]{CarenzaEPL2020}%
  \BibitemOpen
  \bibfield  {author} {\bibinfo {author} {\bibfnamefont {L.N.}\ \bibnamefont
  {Carenza}}, \bibinfo {author} {\bibfnamefont {L.}~\bibnamefont {Biferale}}, \
  and\ \bibinfo {author} {\bibfnamefont {G.}~\bibnamefont {Gonnella}},\
  }\bibfield  {title} {\enquote {\bibinfo {title} {Cascade or not cascade?
  energy transfer and elastic effects in active nematics},}\ }\href@noop {}
  {\bibfield  {journal} {\bibinfo  {journal} {Europhys. Lett.}\ }\textbf
  {\bibinfo {volume} {132}},\ \bibinfo {pages} {44003} (\bibinfo {year}
  {2020}{\natexlab{b}})}\BibitemShut {NoStop}%
\bibitem [{\citenamefont {Alert}\ \emph {et~al.}(2020)\citenamefont {Alert},
  \citenamefont {Joanny},\ and\ \citenamefont {Casademunt}}]{AlertNP2020}%
  \BibitemOpen
  \bibfield  {author} {\bibinfo {author} {\bibfnamefont {R.}~\bibnamefont
  {Alert}}, \bibinfo {author} {\bibfnamefont {J.-F.}\ \bibnamefont {Joanny}}, \
  and\ \bibinfo {author} {\bibfnamefont {J.}~\bibnamefont {Casademunt}},\
  }\bibfield  {title} {\enquote {\bibinfo {title} {Universal scaling of active
  nematic turbulence},}\ }\href@noop {} {\bibfield  {journal} {\bibinfo
  {journal} {Nat. Phys.}\ }\textbf {\bibinfo {volume} {16}},\ \bibinfo {pages}
  {682--688} (\bibinfo {year} {2020})}\BibitemShut {NoStop}%
\bibitem [{\citenamefont {Rorai}\ \emph {et~al.}(2022)\citenamefont {Rorai},
  \citenamefont {Toschi},\ and\ \citenamefont {Pagonabarraga}}]{RoraiPRL2022}%
  \BibitemOpen
  \bibfield  {author} {\bibinfo {author} {\bibfnamefont {C.}~\bibnamefont
  {Rorai}}, \bibinfo {author} {\bibfnamefont {F.}~\bibnamefont {Toschi}}, \
  and\ \bibinfo {author} {\bibfnamefont {I.}~\bibnamefont {Pagonabarraga}},\
  }\bibfield  {title} {\enquote {\bibinfo {title} {Coexistence of active and
  hydrodynamic turbulence in two-dimensional active nematics},}\ }\href@noop {}
  {\bibfield  {journal} {\bibinfo  {journal} {Phys. Rev. Lett.}\ }\textbf
  {\bibinfo {volume} {129}},\ \bibinfo {pages} {218001} (\bibinfo {year}
  {2022})}\BibitemShut {NoStop}%
\bibitem [{\citenamefont {Kruse}\ \emph {et~al.}(2004)\citenamefont {Kruse},
  \citenamefont {Joanny}, \citenamefont {J\"ulicher}, \citenamefont {Prost},\
  and\ \citenamefont {Sekimoto}}]{KrusePRL2004}%
  \BibitemOpen
  \bibfield  {author} {\bibinfo {author} {\bibfnamefont {K.}~\bibnamefont
  {Kruse}}, \bibinfo {author} {\bibfnamefont {J.F.}\ \bibnamefont {Joanny}},
  \bibinfo {author} {\bibfnamefont {F.}~\bibnamefont {J\"ulicher}}, \bibinfo
  {author} {\bibfnamefont {J.}~\bibnamefont {Prost}}, \ and\ \bibinfo {author}
  {\bibfnamefont {K.}~\bibnamefont {Sekimoto}},\ }\bibfield  {title} {\enquote
  {\bibinfo {title} {Asters, vortices and rotating spirals in active gels of
  polar filaments},}\ }\href@noop {} {\bibfield  {journal} {\bibinfo  {journal}
  {Phys. Rev. Lett.}\ }\textbf {\bibinfo {volume} {92}},\ \bibinfo {pages}
  {078101} (\bibinfo {year} {2004})}\BibitemShut {NoStop}%
\bibitem [{\citenamefont {Giomi}\ \emph {et~al.}(2008)\citenamefont {Giomi},
  \citenamefont {Marchetti},\ and\ \citenamefont {Liverpool}}]{GiomiPRL2008}%
  \BibitemOpen
  \bibfield  {author} {\bibinfo {author} {\bibfnamefont {L.}~\bibnamefont
  {Giomi}}, \bibinfo {author} {\bibfnamefont {M.C.}\ \bibnamefont {Marchetti}},
  \ and\ \bibinfo {author} {\bibfnamefont {T.B.}\ \bibnamefont {Liverpool}},\
  }\bibfield  {title} {\enquote {\bibinfo {title} {Complex spontaneous flows
  and concentration banding in active polar films},}\ }\href@noop {} {\bibfield
   {journal} {\bibinfo  {journal} {Phys. Rev. Lett.}\ }\textbf {\bibinfo
  {volume} {101}},\ \bibinfo {pages} {198101} (\bibinfo {year}
  {2008})}\BibitemShut {NoStop}%
\bibitem [{\citenamefont {Edwards}\ and\ \citenamefont
  {Yeomans}(2008)}]{EdwardsEPL2009}%
  \BibitemOpen
  \bibfield  {author} {\bibinfo {author} {\bibfnamefont {S.A.}\ \bibnamefont
  {Edwards}}\ and\ \bibinfo {author} {\bibfnamefont {J.M.}\ \bibnamefont
  {Yeomans}},\ }\bibfield  {title} {\enquote {\bibinfo {title} {Spontaneous
  flow states in active nematics: A unified picture},}\ }\href@noop {}
  {\bibfield  {journal} {\bibinfo  {journal} {Europhys. Lett.}\ }\textbf
  {\bibinfo {volume} {85}},\ \bibinfo {pages} {18008} (\bibinfo {year}
  {2008})}\BibitemShut {NoStop}%
\bibitem [{\citenamefont {Tjhung}\ \emph {et~al.}(2011)\citenamefont {Tjhung},
  \citenamefont {Cates},\ and\ \citenamefont {Marenduzzo}}]{TjhungSM2011}%
  \BibitemOpen
  \bibfield  {author} {\bibinfo {author} {\bibfnamefont {E.}~\bibnamefont
  {Tjhung}}, \bibinfo {author} {\bibfnamefont {M.E.}\ \bibnamefont {Cates}}, \
  and\ \bibinfo {author} {\bibfnamefont {D.}~\bibnamefont {Marenduzzo}},\
  }\bibfield  {title} {\enquote {\bibinfo {title} {Nonequilibrium steady states
  in polar active fluids},}\ }\href@noop {} {\bibfield  {journal} {\bibinfo
  {journal} {Soft Matter}\ }\textbf {\bibinfo {volume} {7}},\ \bibinfo {pages}
  {7453--7464} (\bibinfo {year} {2011})}\BibitemShut {NoStop}%
\bibitem [{\citenamefont {S{\l}omka}\ and\ \citenamefont
  {Dunkel}(2015)}]{SlomkaEPJST2015}%
  \BibitemOpen
  \bibfield  {author} {\bibinfo {author} {\bibfnamefont {L.}~\bibnamefont
  {S{\l}omka}}\ and\ \bibinfo {author} {\bibfnamefont {J.}~\bibnamefont
  {Dunkel}},\ }\bibfield  {title} {\enquote {\bibinfo {title} {Generalized
  navier-stokes equations for active suspensions},}\ }\href@noop {} {\bibfield
  {journal} {\bibinfo  {journal} {Eur. Phys. J. Spec. Top.}\ }\textbf {\bibinfo
  {volume} {224}},\ \bibinfo {pages} {1349--1358} (\bibinfo {year}
  {2015})}\BibitemShut {NoStop}%
\bibitem [{\citenamefont {Stenhammar}\ \emph {et~al.}(2017)\citenamefont
  {Stenhammar}, \citenamefont {Nardini}, , \citenamefont {Nash}, \citenamefont
  {Marenduzzo},\ and\ \citenamefont {Morozov}}]{StenhammarPRL2017}%
  \BibitemOpen
  \bibfield  {author} {\bibinfo {author} {\bibfnamefont {J.}~\bibnamefont
  {Stenhammar}}, \bibinfo {author} {\bibfnamefont {C.}~\bibnamefont {Nardini}},
  , \bibinfo {author} {\bibfnamefont {R.W.}\ \bibnamefont {Nash}}, \bibinfo
  {author} {\bibfnamefont {D.}~\bibnamefont {Marenduzzo}}, \ and\ \bibinfo
  {author} {\bibfnamefont {A.}~\bibnamefont {Morozov}},\ }\bibfield  {title}
  {\enquote {\bibinfo {title} {Role of correlations in the collective behavior
  of microswimmer suspensions},}\ }\href@noop {} {\bibfield  {journal}
  {\bibinfo  {journal} {Phys. Rev. Lett.}\ }\textbf {\bibinfo {volume} {119}},\
  \bibinfo {pages} {028005} (\bibinfo {year} {2017})}\BibitemShut {NoStop}%
\bibitem [{\citenamefont {B\'ardfalvy}\ \emph {et~al.}(2019)\citenamefont
  {B\'ardfalvy}, \citenamefont {Nordanger}, \citenamefont {Nardini},
  \citenamefont {Morozov},\ and\ \citenamefont {Stenhammar}}]{BardfalvySM2019}%
  \BibitemOpen
  \bibfield  {author} {\bibinfo {author} {\bibfnamefont {D.}~\bibnamefont
  {B\'ardfalvy}}, \bibinfo {author} {\bibfnamefont {H.}~\bibnamefont
  {Nordanger}}, \bibinfo {author} {\bibfnamefont {C.}~\bibnamefont {Nardini}},
  \bibinfo {author} {\bibfnamefont {A.}~\bibnamefont {Morozov}}, \ and\
  \bibinfo {author} {\bibfnamefont {J.}~\bibnamefont {Stenhammar}},\ }\bibfield
   {title} {\enquote {\bibinfo {title} {Particle-resolved lattice boltzmann
  simulations of 3-dimensional active turbulence},}\ }\href@noop {} {\bibfield
  {journal} {\bibinfo  {journal} {Soft Matter}\ }\textbf {\bibinfo {volume}
  {15}},\ \bibinfo {pages} {7747--7756} (\bibinfo {year} {2019})}\BibitemShut
  {NoStop}%
\bibitem [{\citenamefont {\v{S}kult\'ety}\ \emph {et~al.}(2020)\citenamefont
  {\v{S}kult\'ety}, \citenamefont {Nardini}, \citenamefont {Stenhammar},
  \citenamefont {Marenduzzo},\ and\ \citenamefont {Morozov}}]{SkultetyPRX2020}%
  \BibitemOpen
  \bibfield  {author} {\bibinfo {author} {\bibfnamefont {V.}~\bibnamefont
  {\v{S}kult\'ety}}, \bibinfo {author} {\bibfnamefont {C.}~\bibnamefont
  {Nardini}}, \bibinfo {author} {\bibfnamefont {J.}~\bibnamefont {Stenhammar}},
  \bibinfo {author} {\bibfnamefont {D.}~\bibnamefont {Marenduzzo}}, \ and\
  \bibinfo {author} {\bibfnamefont {A.}~\bibnamefont {Morozov}},\ }\bibfield
  {title} {\enquote {\bibinfo {title} {Swimming suppresses correlations in
  dilute suspensions of pusher microorganisms},}\ }\href@noop {} {\bibfield
  {journal} {\bibinfo  {journal} {Phys. Rev. X}\ }\textbf {\bibinfo {volume}
  {10}},\ \bibinfo {pages} {031059} (\bibinfo {year} {2020})}\BibitemShut
  {NoStop}%
\bibitem [{\citenamefont {Wu}\ and\ \citenamefont
  {Libchaber}(2000)}]{WuPRL2000}%
  \BibitemOpen
  \bibfield  {author} {\bibinfo {author} {\bibfnamefont {X.-L.}\ \bibnamefont
  {Wu}}\ and\ \bibinfo {author} {\bibfnamefont {A.}~\bibnamefont {Libchaber}},\
  }\bibfield  {title} {\enquote {\bibinfo {title} {Particle diffusion in a
  quasi-two-dimensional bacterial bath},}\ }\href@noop {} {\bibfield  {journal}
  {\bibinfo  {journal} {Phys. Rev. Lett.}\ }\textbf {\bibinfo {volume} {84}},\
  \bibinfo {pages} {3017--3020} (\bibinfo {year} {2000})}\BibitemShut {NoStop}%
\bibitem [{\citenamefont {Hern\'andez-Ortiz}\ \emph {et~al.}(2005)\citenamefont
  {Hern\'andez-Ortiz}, \citenamefont {Stoltz},\ and\ \citenamefont
  {Graham}}]{HernandezOrtizPRL2005}%
  \BibitemOpen
  \bibfield  {author} {\bibinfo {author} {\bibfnamefont {J.P.}\ \bibnamefont
  {Hern\'andez-Ortiz}}, \bibinfo {author} {\bibfnamefont {C.G.}\ \bibnamefont
  {Stoltz}}, \ and\ \bibinfo {author} {\bibfnamefont {M.D.}\ \bibnamefont
  {Graham}},\ }\bibfield  {title} {\enquote {\bibinfo {title} {Transport and
  collective dynamics in suspensions of confined swimming particles},}\
  }\href@noop {} {\bibfield  {journal} {\bibinfo  {journal} {Phys. Rev. Lett.}\
  }\textbf {\bibinfo {volume} {95}},\ \bibinfo {pages} {204501} (\bibinfo
  {year} {2005})}\BibitemShut {NoStop}%
\bibitem [{\citenamefont {Leptos}\ \emph {et~al.}(2009)\citenamefont {Leptos},
  \citenamefont {Guasto}, \citenamefont {Gollub}, \citenamefont {Pesci},\ and\
  \citenamefont {Goldstein}}]{LeptosPRL2009}%
  \BibitemOpen
  \bibfield  {author} {\bibinfo {author} {\bibfnamefont {K.C.}\ \bibnamefont
  {Leptos}}, \bibinfo {author} {\bibfnamefont {J.S.}\ \bibnamefont {Guasto}},
  \bibinfo {author} {\bibfnamefont {J.P.}\ \bibnamefont {Gollub}}, \bibinfo
  {author} {\bibfnamefont {A.I.}\ \bibnamefont {Pesci}}, \ and\ \bibinfo
  {author} {\bibfnamefont {R.E.}\ \bibnamefont {Goldstein}},\ }\bibfield
  {title} {\enquote {\bibinfo {title} {Dynamics of enhanced tracer diffusion in
  suspensions of swimming eukaryotic microorganisms},}\ }\href@noop {}
  {\bibfield  {journal} {\bibinfo  {journal} {Phys. Rev. Lett.}\ }\textbf
  {\bibinfo {volume} {103}},\ \bibinfo {pages} {198103} (\bibinfo {year}
  {2009})}\BibitemShut {NoStop}%
\bibitem [{\citenamefont {Saintillan}\ and\ \citenamefont
  {Shelley}(2012)}]{SaintillianJRSI2012}%
  \BibitemOpen
  \bibfield  {author} {\bibinfo {author} {\bibfnamefont {D.}~\bibnamefont
  {Saintillan}}\ and\ \bibinfo {author} {\bibfnamefont {M.J.}\ \bibnamefont
  {Shelley}},\ }\bibfield  {title} {\enquote {\bibinfo {title} {Emergence of
  coherent structures and large-scale flows in motile suspensions},}\
  }\href@noop {} {\bibfield  {journal} {\bibinfo  {journal} {J. R. Soc.
  Interface}\ }\textbf {\bibinfo {volume} {9}},\ \bibinfo {pages} {571--585}
  (\bibinfo {year} {2012})}\BibitemShut {NoStop}%
\bibitem [{\citenamefont {Biferale}(2004)}]{BiferaleARFM2003}%
  \BibitemOpen
  \bibfield  {author} {\bibinfo {author} {\bibfnamefont {L.}~\bibnamefont
  {Biferale}},\ }\bibfield  {title} {\enquote {\bibinfo {title} {Shell models
  of energy cascade in turbulence},}\ }\href@noop {} {\bibfield  {journal}
  {\bibinfo  {journal} {Annu. Rev. Fluid Mech.}\ }\textbf {\bibinfo {volume}
  {35}},\ \bibinfo {pages} {441--468} (\bibinfo {year} {2004})}\BibitemShut
  {NoStop}%
\bibitem [{\citenamefont {Succi}(2018)}]{Succi}%
  \BibitemOpen
  \bibfield  {author} {\bibinfo {author} {\bibfnamefont {S.}~\bibnamefont
  {Succi}},\ }\href@noop {} {\emph {\bibinfo {title} {The lattice Boltzmann
  equation for complex state of flowing matter}}}\ (\bibinfo  {publisher}
  {Oxford University Press},\ \bibinfo {year} {2018})\BibitemShut {NoStop}%
\bibitem [{\citenamefont {Wolf-Gladow}(2000)}]{WolfGladrow}%
  \BibitemOpen
  \bibfield  {author} {\bibinfo {author} {\bibfnamefont {D.A.}\ \bibnamefont
  {Wolf-Gladow}},\ }\href@noop {} {\emph {\bibinfo {title} {Lattice--gas
  cellular automata and lattice Boltzmann models. An introduction}}}\ (\bibinfo
   {publisher} {Springer},\ \bibinfo {year} {2000})\BibitemShut {NoStop}%
\bibitem [{\citenamefont {Desplat}\ \emph {et~al.}(2001)\citenamefont
  {Desplat}, \citenamefont {Pagonabarraga},\ and\ \citenamefont
  {Bladon}}]{DesplatCPC2001}%
  \BibitemOpen
  \bibfield  {author} {\bibinfo {author} {\bibfnamefont {J.-C.}\ \bibnamefont
  {Desplat}}, \bibinfo {author} {\bibfnamefont {I.}~\bibnamefont
  {Pagonabarraga}}, \ and\ \bibinfo {author} {\bibfnamefont {P.}~\bibnamefont
  {Bladon}},\ }\bibfield  {title} {\enquote {\bibinfo {title} {Ludwig: A
  parallel lattice--boltzmann code for complex fluids},}\ }\href@noop {}
  {\bibfield  {journal} {\bibinfo  {journal} {Comp. Phys. Commun.}\ }\textbf
  {\bibinfo {volume} {134}},\ \bibinfo {pages} {273--290} (\bibinfo {year}
  {2001})}\BibitemShut {NoStop}%
\bibitem [{\citenamefont {Ladd}(1994)}]{LaddJFM1994}%
  \BibitemOpen
  \bibfield  {author} {\bibinfo {author} {\bibfnamefont {A.J.C.}\ \bibnamefont
  {Ladd}},\ }\bibfield  {title} {\enquote {\bibinfo {title} {Numerical
  simulations of particulate suspensions via a discretized boltzmann equation.
  part 1. theoretical foundation},}\ }\href@noop {} {\bibfield  {journal}
  {\bibinfo  {journal} {J. Fluid Mech.}\ }\textbf {\bibinfo {volume} {271}},\
  \bibinfo {pages} {285--309} (\bibinfo {year} {1994})}\BibitemShut {NoStop}%
\bibitem [{\citenamefont {Nguyen}\ and\ \citenamefont
  {Ladd}(2002)}]{NguyenPRE2002}%
  \BibitemOpen
  \bibfield  {author} {\bibinfo {author} {\bibfnamefont {N.Q.}\ \bibnamefont
  {Nguyen}}\ and\ \bibinfo {author} {\bibfnamefont {A.J.C.}\ \bibnamefont
  {Ladd}},\ }\bibfield  {title} {\enquote {\bibinfo {title} {Lubrication
  corrections for lattice-boltzmann simulations of particle suspensions},}\
  }\href@noop {} {\bibfield  {journal} {\bibinfo  {journal} {Phys. Rev. E}\
  }\textbf {\bibinfo {volume} {66}},\ \bibinfo {pages} {046708} (\bibinfo
  {year} {2002})}\BibitemShut {NoStop}%
\bibitem [{\citenamefont {Aidun}\ and\ \citenamefont
  {Clausen}(2010)}]{AidunARFM2010}%
  \BibitemOpen
  \bibfield  {author} {\bibinfo {author} {\bibfnamefont {C.K.}\ \bibnamefont
  {Aidun}}\ and\ \bibinfo {author} {\bibfnamefont {J.R.}\ \bibnamefont
  {Clausen}},\ }\bibfield  {title} {\enquote {\bibinfo {title}
  {Lattice-boltzmann method for complex flows},}\ }\href@noop {} {\bibfield
  {journal} {\bibinfo  {journal} {Annu. Rev. Fluid Mech.}\ }\textbf {\bibinfo
  {volume} {42}},\ \bibinfo {pages} {439--472} (\bibinfo {year}
  {2010})}\BibitemShut {NoStop}%
\bibitem [{\citenamefont {Blake}(1971)}]{BlakeJFM1971}%
  \BibitemOpen
  \bibfield  {author} {\bibinfo {author} {\bibfnamefont {J.R.}\ \bibnamefont
  {Blake}},\ }\bibfield  {title} {\enquote {\bibinfo {title} {A spherical
  envelope approach to ciliary propulsion},}\ }\href@noop {} {\bibfield
  {journal} {\bibinfo  {journal} {J. Fluid Mech.}\ }\textbf {\bibinfo {volume}
  {46}},\ \bibinfo {pages} {199--208} (\bibinfo {year} {1971})}\BibitemShut
  {NoStop}%
\bibitem [{\citenamefont {Ishikawa}\ \emph {et~al.}(2006)\citenamefont
  {Ishikawa}, \citenamefont {Simmonds},\ and\ \citenamefont
  {Pedley}}]{IshikawaJFM2006}%
  \BibitemOpen
  \bibfield  {author} {\bibinfo {author} {\bibfnamefont {T.}~\bibnamefont
  {Ishikawa}}, \bibinfo {author} {\bibfnamefont {M.P.}\ \bibnamefont
  {Simmonds}}, \ and\ \bibinfo {author} {\bibfnamefont {T.J.}\ \bibnamefont
  {Pedley}},\ }\href@noop {} {\bibfield  {journal} {\bibinfo  {journal} {J.
  Fluid Mech.}\ }\textbf {\bibinfo {volume} {568}},\ \bibinfo {pages}
  {119--160} (\bibinfo {year} {2006})}\BibitemShut {NoStop}%
\bibitem [{\citenamefont {Matas~Navarro}\ and\ \citenamefont
  {Pagonabarraga}(2010)}]{MatasNavarroEPJE2010}%
  \BibitemOpen
  \bibfield  {author} {\bibinfo {author} {\bibfnamefont {R.}~\bibnamefont
  {Matas~Navarro}}\ and\ \bibinfo {author} {\bibfnamefont {I.}~\bibnamefont
  {Pagonabarraga}},\ }\href@noop {} {\bibfield  {journal} {\bibinfo  {journal}
  {Eur. Phys. J. E}\ }\textbf {\bibinfo {volume} {33}},\ \bibinfo {pages}
  {27--39} (\bibinfo {year} {2010})}\BibitemShut {NoStop}%
\bibitem [{\citenamefont {Alarc\'on}\ and\ \citenamefont
  {Pagonabarraga}(2013)}]{AlarconJML2013}%
  \BibitemOpen
  \bibfield  {author} {\bibinfo {author} {\bibfnamefont {F.}~\bibnamefont
  {Alarc\'on}}\ and\ \bibinfo {author} {\bibfnamefont {I.}~\bibnamefont
  {Pagonabarraga}},\ }\bibfield  {title} {\enquote {\bibinfo {title}
  {Spontaneous aggregation and global polar ordering in squirmer
  suspensions},}\ }\href@noop {} {\bibfield  {journal} {\bibinfo  {journal} {J.
  Mol. Liq.}\ }\textbf {\bibinfo {volume} {185}},\ \bibinfo {pages} {56--61}
  (\bibinfo {year} {2013})}\BibitemShut {NoStop}%
\bibitem [{\citenamefont {Alarc\'on}\ \emph {et~al.}(2017)\citenamefont
  {Alarc\'on}, \citenamefont {Valeriani},\ and\ \citenamefont
  {Pagonabarraga}}]{AlarconSM2017}%
  \BibitemOpen
  \bibfield  {author} {\bibinfo {author} {\bibfnamefont {F.}~\bibnamefont
  {Alarc\'on}}, \bibinfo {author} {\bibfnamefont {C.}~\bibnamefont
  {Valeriani}}, \ and\ \bibinfo {author} {\bibfnamefont {I.}~\bibnamefont
  {Pagonabarraga}},\ }\bibfield  {title} {\enquote {\bibinfo {title}
  {Morphology of clusters of attractive dry and wet self-propelled spherical
  particle suspensions},}\ }\href@noop {} {\bibfield  {journal} {\bibinfo
  {journal} {Soft Matter}\ }\textbf {\bibinfo {volume} {13}},\ \bibinfo {pages}
  {814--826} (\bibinfo {year} {2017})}\BibitemShut {NoStop}%
\bibitem [{\citenamefont {Scagliarini}\ and\ \citenamefont
  {Pagonabarraga}(2022)}]{ScagliariniSM2022}%
  \BibitemOpen
  \bibfield  {author} {\bibinfo {author} {\bibfnamefont {A.}~\bibnamefont
  {Scagliarini}}\ and\ \bibinfo {author} {\bibfnamefont {I.}~\bibnamefont
  {Pagonabarraga}},\ }\bibfield  {title} {\enquote {\bibinfo {title}
  {Hydrodynamic and geometric effects in the sedimentation of model
  run-and-tumble microswimmers},}\ }\href@noop {} {\bibfield  {journal}
  {\bibinfo  {journal} {Soft Matter}\ }\textbf {\bibinfo {volume} {18}},\
  \bibinfo {pages} {2407--2413} (\bibinfo {year} {2022})}\BibitemShut {NoStop}%
\bibitem [{\citenamefont {Evans}\ \emph {et~al.}(2011)\citenamefont {Evans},
  \citenamefont {Ishikawa}, \citenamefont {Yamaguchi},\ and\ \citenamefont
  {Lauga}}]{EvansPOF2011}%
  \BibitemOpen
  \bibfield  {author} {\bibinfo {author} {\bibfnamefont {A.A.}\ \bibnamefont
  {Evans}}, \bibinfo {author} {\bibfnamefont {T}~\bibnamefont {Ishikawa}},
  \bibinfo {author} {\bibfnamefont {T.}~\bibnamefont {Yamaguchi}}, \ and\
  \bibinfo {author} {\bibfnamefont {E.}~\bibnamefont {Lauga}},\ }\bibfield
  {title} {\enquote {\bibinfo {title} {Orientational order in concentrated
  suspensions of spherical microswimmers},}\ }\href@noop {} {\bibfield
  {journal} {\bibinfo  {journal} {Phys. Fluids}\ }\textbf {\bibinfo {volume}
  {23}},\ \bibinfo {pages} {111702} (\bibinfo {year} {2011})}\BibitemShut
  {NoStop}%
\bibitem [{\citenamefont {Liu}\ \emph {et~al.}(2021)\citenamefont {Liu},
  \citenamefont {Zeng}, \citenamefont {Ma},\ and\ \citenamefont
  {Cheng}}]{LiuSM2021}%
  \BibitemOpen
  \bibfield  {author} {\bibinfo {author} {\bibfnamefont {Z.}~\bibnamefont
  {Liu}}, \bibinfo {author} {\bibfnamefont {W.}~\bibnamefont {Zeng}}, \bibinfo
  {author} {\bibfnamefont {X.}~\bibnamefont {Ma}}, \ and\ \bibinfo {author}
  {\bibfnamefont {X.}~\bibnamefont {Cheng}},\ }\bibfield  {title} {\enquote
  {\bibinfo {title} {Density fluctuations and energy spectra of 3d bacterial
  suspensions},}\ }\href@noop {} {\bibfield  {journal} {\bibinfo  {journal}
  {Soft Matter}\ }\textbf {\bibinfo {volume} {17}},\ \bibinfo {pages}
  {10806--10817} (\bibinfo {year} {2021})}\BibitemShut {NoStop}%
\bibitem [{\citenamefont {Lin}\ \emph {et~al.}(2021)\citenamefont {Lin},
  \citenamefont {Zhang}, \citenamefont {Bi}, \citenamefont {Li},\ and\
  \citenamefont {Feng}}]{lin2021}%
  \BibitemOpen
  \bibfield  {author} {\bibinfo {author} {\bibfnamefont {S.-Z.}\ \bibnamefont
  {Lin}}, \bibinfo {author} {\bibfnamefont {W.-Y.}\ \bibnamefont {Zhang}},
  \bibinfo {author} {\bibfnamefont {D.}~\bibnamefont {Bi}}, \bibinfo {author}
  {\bibfnamefont {B.}~\bibnamefont {Li}}, \ and\ \bibinfo {author}
  {\bibfnamefont {X.-Q.}\ \bibnamefont {Feng}},\ }\bibfield  {title} {\enquote
  {\bibinfo {title} {Energetics of mesoscale cell turbulence in two-dimensional
  monolayers},}\ }\href@noop {} {\bibfield  {journal} {\bibinfo  {journal}
  {Commun. Phys.}\ }\textbf {\bibinfo {volume} {4}},\ \bibinfo {pages} {21}
  (\bibinfo {year} {2021})}\BibitemShut {NoStop}%
\bibitem [{\citenamefont {Zaid}\ \emph {et~al.}(2011)\citenamefont {Zaid},
  \citenamefont {Dunkel},\ and\ \citenamefont {Yeomans}}]{ZaidJRSI2011}%
  \BibitemOpen
  \bibfield  {author} {\bibinfo {author} {\bibfnamefont {I.M.}\ \bibnamefont
  {Zaid}}, \bibinfo {author} {\bibfnamefont {J.}~\bibnamefont {Dunkel}}, \ and\
  \bibinfo {author} {\bibfnamefont {J.M.}\ \bibnamefont {Yeomans}},\ }\bibfield
   {title} {\enquote {\bibinfo {title} {L\'evy fluctuations and mixing in
  dilute suspensions of algae and bacteria},}\ }\href@noop {} {\bibfield
  {journal} {\bibinfo  {journal} {J. R. Soc. Interface}\ }\textbf {\bibinfo
  {volume} {8}},\ \bibinfo {pages} {1314--1331} (\bibinfo {year}
  {2011})}\BibitemShut {NoStop}%
\bibitem [{\citenamefont {Rushkin}\ \emph {et~al.}(2010)\citenamefont
  {Rushkin}, \citenamefont {Kantsler},\ and\ \citenamefont
  {Goldstein}}]{RushkinPRL2010}%
  \BibitemOpen
  \bibfield  {author} {\bibinfo {author} {\bibfnamefont {I.}~\bibnamefont
  {Rushkin}}, \bibinfo {author} {\bibfnamefont {V.}~\bibnamefont {Kantsler}}, \
  and\ \bibinfo {author} {\bibfnamefont {R.E.}\ \bibnamefont {Goldstein}},\
  }\bibfield  {title} {\enquote {\bibinfo {title} {Fluid velocity fluctuations
  in a suspension of swimming protists},}\ }\href@noop {} {\bibfield  {journal}
  {\bibinfo  {journal} {Phys. Rev. Lett.}\ }\textbf {\bibinfo {volume} {105}},\
  \bibinfo {pages} {188101} (\bibinfo {year} {2010})}\BibitemShut {NoStop}%
\bibitem [{\citenamefont {Gledzer}(1973)}]{GledzerSPD1973}%
  \BibitemOpen
  \bibfield  {author} {\bibinfo {author} {\bibfnamefont {E.B.}\ \bibnamefont
  {Gledzer}},\ }\bibfield  {title} {\enquote {\bibinfo {title} {System of
  hydrodynamic type admitting two quadratic integrals of motion},}\ }\href@noop
  {} {\bibfield  {journal} {\bibinfo  {journal} {Sov. Phys. Dokl.}\ }\textbf
  {\bibinfo {volume} {18}},\ \bibinfo {pages} {216--217} (\bibinfo {year}
  {1973})}\BibitemShut {NoStop}%
\bibitem [{\citenamefont {Desnyansky}\ and\ \citenamefont
  {Novikov}(1974)}]{DesnyanskyIANSFAO1974}%
  \BibitemOpen
  \bibfield  {author} {\bibinfo {author} {\bibfnamefont {V.N.}\ \bibnamefont
  {Desnyansky}}\ and\ \bibinfo {author} {\bibfnamefont {E.A.}\ \bibnamefont
  {Novikov}},\ }\bibfield  {title} {\enquote {\bibinfo {title} {The evolution
  of turbulence spectra to the similarity regime},}\ }\href@noop {} {\bibfield
  {journal} {\bibinfo  {journal} {Izv. Akad. Nauk SSSR Fiz. Atmos. Okeana}\
  }\textbf {\bibinfo {volume} {10}},\ \bibinfo {pages} {127--136} (\bibinfo
  {year} {1974})}\BibitemShut {NoStop}%
\bibitem [{\citenamefont {Bohr}\ \emph {et~al.}(1998)\citenamefont {Bohr},
  \citenamefont {Jensen}, \citenamefont {Paladin},\ and\ \citenamefont
  {Vulpiani}}]{Bohr}%
  \BibitemOpen
  \bibfield  {author} {\bibinfo {author} {\bibfnamefont {T.}~\bibnamefont
  {Bohr}}, \bibinfo {author} {\bibfnamefont {M.H.}\ \bibnamefont {Jensen}},
  \bibinfo {author} {\bibfnamefont {G.}~\bibnamefont {Paladin}}, \ and\
  \bibinfo {author} {\bibfnamefont {A.}~\bibnamefont {Vulpiani}},\ }\href@noop
  {} {\emph {\bibinfo {title} {Dynamical systems approach to turbulence}}}\
  (\bibinfo  {publisher} {Cambridge University Press},\ \bibinfo {year}
  {1998})\BibitemShut {NoStop}%
\bibitem [{\citenamefont {Benzi}\ \emph {et~al.}(2003)\citenamefont {Benzi},
  \citenamefont {De~Angelis}, \citenamefont {Govindarajan},\ and\ \citenamefont
  {Procaccia}}]{BenziPRE2003}%
  \BibitemOpen
  \bibfield  {author} {\bibinfo {author} {\bibfnamefont {R.}~\bibnamefont
  {Benzi}}, \bibinfo {author} {\bibfnamefont {E.}~\bibnamefont {De~Angelis}},
  \bibinfo {author} {\bibfnamefont {R.}~\bibnamefont {Govindarajan}}, \ and\
  \bibinfo {author} {\bibfnamefont {I.}~\bibnamefont {Procaccia}},\ }\bibfield
  {title} {\enquote {\bibinfo {title} {Shell model of drag reduction with
  polymer additives in homogeneous turbulence},}\ }\href@noop {} {\bibfield
  {journal} {\bibinfo  {journal} {Phys. Rev. E}\ }\textbf {\bibinfo {volume}
  {68}},\ \bibinfo {pages} {016308} (\bibinfo {year} {2003})}\BibitemShut
  {NoStop}%
\bibitem [{\citenamefont {Benzi}\ \emph {et~al.}(2004)\citenamefont {Benzi},
  \citenamefont {Horesh},\ and\ \citenamefont {Procaccia}}]{BenziEPL2004}%
  \BibitemOpen
  \bibfield  {author} {\bibinfo {author} {\bibfnamefont {R.}~\bibnamefont
  {Benzi}}, \bibinfo {author} {\bibfnamefont {N.}~\bibnamefont {Horesh}}, \
  and\ \bibinfo {author} {\bibfnamefont {I.}~\bibnamefont {Procaccia}},\
  }\bibfield  {title} {\enquote {\bibinfo {title} {Shell model of
  two-dimensional turbulence in polymer solutions},}\ }\href@noop {} {\bibfield
   {journal} {\bibinfo  {journal} {Europhys. Lett.}\ }\textbf {\bibinfo
  {volume} {68}},\ \bibinfo {pages} {310--315} (\bibinfo {year}
  {2004})}\BibitemShut {NoStop}%
\bibitem [{\citenamefont {Ray}\ and\ \citenamefont
  {Vincenzi}(2016)}]{RayEPL2016}%
  \BibitemOpen
  \bibfield  {author} {\bibinfo {author} {\bibfnamefont {S.S.}\ \bibnamefont
  {Ray}}\ and\ \bibinfo {author} {\bibfnamefont {D.}~\bibnamefont {Vincenzi}},\
  }\bibfield  {title} {\enquote {\bibinfo {title} {Elastic turbulence in a
  shell model of polymer solution},}\ }\href@noop {} {\bibfield  {journal}
  {\bibinfo  {journal} {Europhys. Lett.}\ }\textbf {\bibinfo {volume} {144}},\
  \bibinfo {pages} {44001} (\bibinfo {year} {2016})}\BibitemShut {NoStop}%
\bibitem [{Note1()}]{Note1}%
  \BibitemOpen
  \bibinfo {note} {The system for the velocity variables is coupled, in this
  case, to a set of equations for analogous {\protect \it polymer} variables,
  interpreted as spectral amplitudes of an auxiliary vector field whose dyadic
  product is the polymer conformation tensor~\cite {BenziPRE2003}.}\BibitemShut
  {Stop}%
\bibitem [{\citenamefont {Yoshinaga}\ and\ \citenamefont
  {Liverpool}(2017)}]{YoshinagaPRE2017}%
  \BibitemOpen
  \bibfield  {author} {\bibinfo {author} {\bibfnamefont {N.}~\bibnamefont
  {Yoshinaga}}\ and\ \bibinfo {author} {\bibfnamefont {T.B.}\ \bibnamefont
  {Liverpool}},\ }\bibfield  {title} {\enquote {\bibinfo {title} {Hydrodynamic
  interactions in dense active suspensions: From polar order to dynamical
  clusters},}\ }\href@noop {} {\bibfield  {journal} {\bibinfo  {journal} {Phys.
  Rev. E}\ }\textbf {\bibinfo {volume} {96}},\ \bibinfo {pages} {020603(R)}
  (\bibinfo {year} {2017})}\BibitemShut {NoStop}%
\bibitem [{\citenamefont {Aditi~Simha}\ and\ \citenamefont
  {Ramaswamy}(2002)}]{SimhaPRL2002}%
  \BibitemOpen
  \bibfield  {author} {\bibinfo {author} {\bibfnamefont {R.}~\bibnamefont
  {Aditi~Simha}}\ and\ \bibinfo {author} {\bibfnamefont {S.}~\bibnamefont
  {Ramaswamy}},\ }\bibfield  {title} {\enquote {\bibinfo {title} {Hydrodynamic
  fluctuations and instabilities in ordered suspensions of self-propelled
  particles},}\ }\href@noop {} {\bibfield  {journal} {\bibinfo  {journal}
  {Phys. Rev. Lett.}\ }\textbf {\bibinfo {volume} {89}},\ \bibinfo {pages}
  {058101} (\bibinfo {year} {2002})}\BibitemShut {NoStop}%
\bibitem [{\citenamefont {Hatwalne}\ \emph {et~al.}(2004)\citenamefont
  {Hatwalne}, \citenamefont {Ramaswamy}, \citenamefont {Rao},\ and\
  \citenamefont {Aditi~Simha}}]{HatwalnePRL2004}%
  \BibitemOpen
  \bibfield  {author} {\bibinfo {author} {\bibfnamefont {Y.}~\bibnamefont
  {Hatwalne}}, \bibinfo {author} {\bibfnamefont {S.}~\bibnamefont {Ramaswamy}},
  \bibinfo {author} {\bibfnamefont {M.}~\bibnamefont {Rao}}, \ and\ \bibinfo
  {author} {\bibfnamefont {R.}~\bibnamefont {Aditi~Simha}},\ }\bibfield
  {title} {\enquote {\bibinfo {title} {Rheology of active-particle
  suspensions},}\ }\href@noop {} {\bibfield  {journal} {\bibinfo  {journal}
  {Phys. Rev. Lett.}\ }\textbf {\bibinfo {volume} {92}},\ \bibinfo {pages}
  {118101} (\bibinfo {year} {2004})}\BibitemShut {NoStop}%
\bibitem [{\citenamefont {Saintillan}\ and\ \citenamefont
  {Shelley}(2008{\natexlab{a}})}]{SaintillianPRL2008}%
  \BibitemOpen
  \bibfield  {author} {\bibinfo {author} {\bibfnamefont {D.}~\bibnamefont
  {Saintillan}}\ and\ \bibinfo {author} {\bibfnamefont {M.J.}\ \bibnamefont
  {Shelley}},\ }\bibfield  {title} {\enquote {\bibinfo {title} {Instabilities
  and pattern formation in active particle suspensions: kinetic theory and
  continuum simulations},}\ }\href@noop {} {\bibfield  {journal} {\bibinfo
  {journal} {Phys. Rev. Lett.}\ }\textbf {\bibinfo {volume} {100}},\ \bibinfo
  {pages} {178103} (\bibinfo {year} {2008}{\natexlab{a}})}\BibitemShut
  {NoStop}%
\bibitem [{\citenamefont {Saintillan}\ and\ \citenamefont
  {Shelley}(2008{\natexlab{b}})}]{SaintillianPOF2008}%
  \BibitemOpen
  \bibfield  {author} {\bibinfo {author} {\bibfnamefont {D.}~\bibnamefont
  {Saintillan}}\ and\ \bibinfo {author} {\bibfnamefont {M.J.}\ \bibnamefont
  {Shelley}},\ }\bibfield  {title} {\enquote {\bibinfo {title} {Instabilities,
  pattern formation, and mixing in active suspensions},}\ }\href@noop {}
  {\bibfield  {journal} {\bibinfo  {journal} {Phys. Fluids}\ }\textbf {\bibinfo
  {volume} {20}},\ \bibinfo {pages} {123304} (\bibinfo {year}
  {2008}{\natexlab{b}})}\BibitemShut {NoStop}%
\bibitem [{\citenamefont {Baskaran}\ and\ \citenamefont
  {Marchetti}(2009)}]{BaskaranPNAS2009}%
  \BibitemOpen
  \bibfield  {author} {\bibinfo {author} {\bibfnamefont {A.}~\bibnamefont
  {Baskaran}}\ and\ \bibinfo {author} {\bibfnamefont {M.C.}\ \bibnamefont
  {Marchetti}},\ }\bibfield  {title} {\enquote {\bibinfo {title} {Statistical
  mechanics and hydrodynamics of bacterial suspensions},}\ }\href@noop {}
  {\bibfield  {journal} {\bibinfo  {journal} {Proc. Natl. Acad. Sci. USA}\
  }\textbf {\bibinfo {volume} {106}},\ \bibinfo {pages} {15567--15572}
  (\bibinfo {year} {2009})}\BibitemShut {NoStop}%
\bibitem [{\citenamefont {Bird}\ \emph {et~al.}(1987)\citenamefont {Bird},
  \citenamefont {Armstrong},\ and\ \citenamefont {Hassager}}]{Bird}%
  \BibitemOpen
  \bibfield  {author} {\bibinfo {author} {\bibfnamefont {R.B.}\ \bibnamefont
  {Bird}}, \bibinfo {author} {\bibfnamefont {R.C.}\ \bibnamefont {Armstrong}},
  \ and\ \bibinfo {author} {\bibfnamefont {O.}~\bibnamefont {Hassager}},\
  }\href@noop {} {\emph {\bibinfo {title} {Dynamics of Polymeric Liquids}}}\
  (\bibinfo  {publisher} {John Wiley and sons},\ \bibinfo {year}
  {1987})\BibitemShut {NoStop}%
\bibitem [{\citenamefont {L'vov}\ \emph {et~al.}(1998)\citenamefont {L'vov},
  \citenamefont {Podivilov}, \citenamefont {Pomyalov}, \citenamefont
  {Procaccia},\ and\ \citenamefont {Vandembroucq}}]{LvovPRE1998}%
  \BibitemOpen
  \bibfield  {author} {\bibinfo {author} {\bibfnamefont {V.S.}\ \bibnamefont
  {L'vov}}, \bibinfo {author} {\bibfnamefont {E.}~\bibnamefont {Podivilov}},
  \bibinfo {author} {\bibfnamefont {A.}~\bibnamefont {Pomyalov}}, \bibinfo
  {author} {\bibfnamefont {I.}~\bibnamefont {Procaccia}}, \ and\ \bibinfo
  {author} {\bibfnamefont {D.}~\bibnamefont {Vandembroucq}},\ }\bibfield
  {title} {\enquote {\bibinfo {title} {Improved shell model of turbulence},}\
  }\href@noop {} {\bibfield  {journal} {\bibinfo  {journal} {Phys. Rev. E}\
  }\textbf {\bibinfo {volume} {58}},\ \bibinfo {pages} {1811--1822} (\bibinfo
  {year} {1998})}\BibitemShut {NoStop}%
\bibitem [{\citenamefont {Gilbert}\ \emph {et~al.}(2002)\citenamefont
  {Gilbert}, \citenamefont {L'vov}, \citenamefont {Pomyalov},\ and\
  \citenamefont {Procaccia}}]{GilbertPRL2002}%
  \BibitemOpen
  \bibfield  {author} {\bibinfo {author} {\bibfnamefont {T.}~\bibnamefont
  {Gilbert}}, \bibinfo {author} {\bibfnamefont {V.S.}\ \bibnamefont {L'vov}},
  \bibinfo {author} {\bibfnamefont {A.}~\bibnamefont {Pomyalov}}, \ and\
  \bibinfo {author} {\bibfnamefont {I.}~\bibnamefont {Procaccia}},\ }\bibfield
  {title} {\enquote {\bibinfo {title} {Inverse cascade regime in shell models
  of two--dimensional turbulence},}\ }\href@noop {} {\bibfield  {journal}
  {\bibinfo  {journal} {Phys. Rev. Lett.}\ }\textbf {\bibinfo {volume} {89}},\
  \bibinfo {pages} {074501} (\bibinfo {year} {2002})}\BibitemShut {NoStop}%
\bibitem [{\citenamefont {Saintillan}\ and\ \citenamefont
  {Shelley}(2007)}]{SaintillianPRL2007}%
  \BibitemOpen
  \bibfield  {author} {\bibinfo {author} {\bibfnamefont {D.}~\bibnamefont
  {Saintillan}}\ and\ \bibinfo {author} {\bibfnamefont {M.J.}\ \bibnamefont
  {Shelley}},\ }\bibfield  {title} {\enquote {\bibinfo {title} {Orientational
  order and instabilities in suspensions of self--locomoting rods},}\
  }\href@noop {} {\bibfield  {journal} {\bibinfo  {journal} {Phys. Rev. Lett.}\
  }\textbf {\bibinfo {volume} {99}},\ \bibinfo {pages} {058102} (\bibinfo
  {year} {2007})}\BibitemShut {NoStop}%
\bibitem [{\citenamefont {Pisarenko}\ \emph {et~al.}(1993)\citenamefont
  {Pisarenko}, \citenamefont {Biferale}, \citenamefont {Courvoisier},
  \citenamefont {Frisch},\ and\ \citenamefont {Vergassola}}]{PisarenkoPOF1993}%
  \BibitemOpen
  \bibfield  {author} {\bibinfo {author} {\bibfnamefont {D.}~\bibnamefont
  {Pisarenko}}, \bibinfo {author} {\bibfnamefont {L.}~\bibnamefont {Biferale}},
  \bibinfo {author} {\bibfnamefont {D.}~\bibnamefont {Courvoisier}}, \bibinfo
  {author} {\bibfnamefont {U.}~\bibnamefont {Frisch}}, \ and\ \bibinfo {author}
  {\bibfnamefont {M.}~\bibnamefont {Vergassola}},\ }\bibfield  {title}
  {\enquote {\bibinfo {title} {Further results on multifractality in shell
  models},}\ }\href@noop {} {\bibfield  {journal} {\bibinfo  {journal} {Phys.
  Fluids}\ }\textbf {\bibinfo {volume} {5}},\ \bibinfo {pages} {2533--2538}
  (\bibinfo {year} {1993})}\BibitemShut {NoStop}%
\end{thebibliography}%


\end{document}